\documentclass[aps,prx,reprint,floatfix,longbibliography]{revtex4-1}

\usepackage{amsmath,amssymb,tikz,bbm}

\newcommand{\ie}{{\it i.e.}}
\newcommand{\eg}{{\it e.g.}}

\newcommand{\bra}[1]{\left\langle#1\right|}
\newcommand{\ket}[1]{\left|#1\right\rangle}

\newcommand{\ig}{\text{i}}
\DeclareMathOperator{\tr}{tr}
\newcommand{\spin}[2]{\hat{S}^{#1}_{#2}}
\newcommand{\TT}{^1(\text{T}\cdots\text{T})}
\newcommand{\ex}{\text{e}}
\newcommand{\viz}{\emph{viz.}}

\usepackage{graphicx}
\usepackage{bm}

\begin{document}

\widetext

\title{Triplet-Triplet Decoherence in Singlet Fission}

\author{Max Marcus}
\email{max.marcus@chem.ox.ac.uk}
\affiliation{Department of Chemistry, Physical \& Theoretical Chemistry Laboratory, University of Oxford, Oxford, OX1 3QZ, United Kingdom.}
\author{William Barford}
\email{william.barford@chem.ox.ac.uk}
\affiliation{Department of Chemistry, Physical \& Theoretical Chemistry Laboratory, University of Oxford, Oxford, OX1 3QZ, United Kingdom.}

\begin{abstract}
Singlet fission is commonly defined to involve a process by which an overall singlet state with local triplet structure spin-decoheres into two triplet states, thereby completing the fission process. This process, often defined in loose terms involving the multiplicity of the overall state, is investigated here using a uniform Heisenberg spin-chain subject to a dephasing environmental interaction. We introduce new results from quantum information theory which enables the quantification of coherence and entanglement in a bi- and multipartite system. The calculated measures of these quantum effects can be linked to observables, such as magnetisation and total spin, with simulations of the model and using theoretical methods. We demonstrate that these observables can act as a proxy for the coherence and entanglement measures. The decay of both of these between the two local triplets can be monitored, enabling a clear definition of the spin-decoherence process in singlet fission.
\end{abstract}

\maketitle

\section{Introduction}
Singlet fission (SF) is a process which has received increased interest with the possibility to be utilised in a wide range of applications\cite{Smith:2010aa,Jadhav:2012aa,Congreve:2013aa,Musser:2019aa}. For instance, the creation of two electronic excitations by absorbing one photon could lead to an increase in the efficiency of photovoltaics beyond the Schockley-Queisser limit\cite{Nelson:2013aa,Rao:2017aa}. The application of a transparent layer of SF-capable material onto a solar cell and the subsequent harnessing of high-energy photons has the potential to increase the efficiency of even the best currently available cells. As such, understanding the mechanism of SF is of great importance in order to be able to tune materials to desired properties. 

Singlet fission, especially in carotenoids, has also been observed in nature. For instance, in light-absorbing complexes of bacteria, carotenoids have been found to absorb in the blue-green region of the visible spectrum, complementing the absorption of longer wavelengths by the chlorophyll complexes. Simultaneously, they act as triplet quenchers for the chlorophyll, preventing the formation of singlet dioxygen in living cells\cite{Peterman:1995aa,Ritz:2000aa,Polivka:2004aa,Polivka:2010aa}.

\begin{figure}
\centering
\includegraphics[width = 0.75\linewidth]{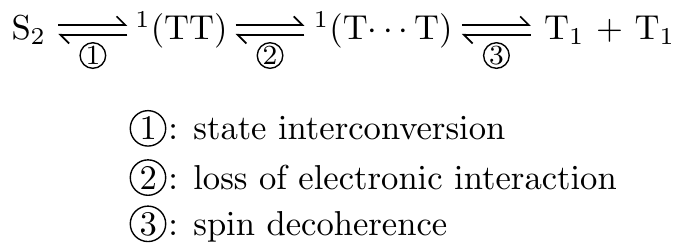}
\caption{\label{Fig:1}Proposed scheme for singlet fission. Note that step 1 is not relevant for polyenes \cite{Valentine:2020aa}. The initially excited singlet state interconverts (if necessary) into a singlet state with local triplet character. Via electronic decoherence the triplets dissociate forming a geminate triplet pair, which subsequently decoheres into two truly independent triplets. The proposed time scale for the third step is on a different order of magnitude than the first two.}
\end{figure}

Although ubiquitous in nature and likely useful in technology, little is known about the actual process of SF. Several schemes of the singlet decay have been proposed, and the most commonly encountered and currently favoured picture is shown in Fig. \ref{Fig:1}\cite{Casanova:2018aa,Scholes:2015aa,Miyata:2019aa}. The debate in the current literature especially involves the second and third steps, which have been characterised as first losing electronic interaction, mainly by the triplet states migrating away from each other, and second the loss of spin coherence and the emergence of overall independent triplet states, \ie\ non-geminate triplet pairs. Steps 1 and 2, if they can be characterised as above, are spin-allowed processes and are therefore expected to happen on a faster time scale than the third step, which necessarily involves an interaction of the spins with an external field. This interaction can be with either nuclear spin, other electrons, or other local or global magnetic fields, which allow for an acquisition of overall spin in the third step and, in particular, for $\left\langle S^2\right\rangle \neq 0$.

Experimentally it has been observed in $\pi$-conjugated polymers, \eg, polydiacetylene \cite{Pandaya:2020aa} and oligo(thienylene-vinylenes) \cite{Musser:2019ab}, that an optical excitation above the $S_2$ (\ie\ $1^1B_u^+$) manifold band edge leads to the creation of non-geminate pairs of triplets. These non-geminate pairs are observed on a timescale of tens of nanoseconds \cite{Rao:2017aa}. The precise mechanism for this process in polymers is not fully understood, but intermediate highly-correlated singlet states with significant triplet pair (or bimagnon) character are believed to participate. Candidate states are the $2^1A_g^-$ ($S_1$) state, and the slightly higher in energy $1^1B_u^-$ and $3^1A_g^-$ states.

Recent theoretical modelling by Valentine, Manawadu and Barford \cite{Valentine:2020aa} of the Pariser-Parr-Pople-Peierls model of $\pi$-conjugated electrons has identified the $3^1A_g^-$ state as a viable candidate for this intermediate state. As shown in \cite{Valentine:2020aa}, the $3^1A_g^-$ state corresponds to a pair of unbound triplets (in contrast to the $2^1A_g^-$ state, which is described as a pair of bound triplet states \cite{Tavan:1987aa,Barford:2001aa,Schmidt:2012aa,Barford:2013ab}). The $3^1A_g^-$ state may thus be labelled as $\TT$, meaning that it is a pair of electronically uncoupled, but quantum mechanically entangled pair of triplets (\ie, a geminate pair). It is the loss of entanglement of this triplet pair that we describe in this paper.

To do so, we will consider the last step of the scheme in Fig. \ref{Fig:1}, the spin decoherence, and just assume that the system has undergone the first two steps in some way (if applicable), leading to a pair of paired triplets with overall singlet multiplicity, so-called geminate triplets. Given the nature of this triplet pair state and the spin pairing within it, it will have a `memory' of its previous state - the triplets making up the $\TT$ state are not independent. The loss of this information connecting the triplets is what we define as singlet fission, in particular the loss of spin coherence. The questions then arise, how can we define the loss of spin coherence and motivate such a definition theoretically, and how can it be observed.

In section II we, therefore, introduce the Hamiltonian used to model a uniform spin chain, including the basis set employed to investigate singlet fission. This is followed by the discussion of environmental effects, incorporated using a Lindblad master equation, by which the steady states are investigated. We will show that equations of motion can be derived for observables, as well as for quantum measures on entanglement and coherence, and that these show the same decay rate. In section IV we will then compare the theoretical predictions to numerical simulations and show that we can use observables as proxies to investigate quantum effects.

\section{Model Hamiltonian}
\subsection{Uniform Heisenberg Chain}
As described in the Introduction, the candidate $\TT$ state in $\pi$-conjugated polymers is the $3^1A_g^-$ state: an electronically uncoupled but spin-correlated triplet pair (or bimagnon). This state is almost entirely covalent in character, \ie, it has negligible electron-hole character \cite{Valentine:2020aa}. As such, it is accurately described by the Heisenberg model of antiferromagnets, which may formally be derived in the large $U$ limit of the Pariser-Parr-Pople model \cite{Barford:2013ab}. The uniform antiferromagnetic Heisenberg model is given as,
\begin{equation}
\hat{H}_0 = J\sum_{i=1}^{N-1}\hat{\bf S}_i\cdot\hat{\bf S}_{i+1},
\label{Eq:HXXX}
\end{equation}
where $\hat{\bf S}_i$ is the spin operator of spin $i$ coupled to its nearest neighbours with coupling strength $J>0$. We keep the model intentionally simple to ease solving the relevant equations. However, it is straightforward to introduce additional terms, such as dipolar coupling, or dimerize the chain to model alternating bond lengths.

The basis used to expand the Hamiltonian is a local spin basis with states $\left\{\ket{\bm \sigma_i}\right\}$ where $\ket{\bm\sigma_i} = \ket{\sigma_1\sigma_2\ldots\sigma_N}$ and each $\sigma$ denotes the eigenvalue of the $\spin{z}{i}$ operator, which can take the values $\pm \hbar/2$, corresponding to the states $\ket{\uparrow}$ and $\ket{\downarrow}$ for each site. The groundstate of this model is generally a singlet state if $J>0$, giving rise to the correct anti-ferromagnetic behaviour.  We will discuss the basis set used for the chains in more detail below.

As the spins considered in this work are located in $\pi$-orbitals which have non-zero orbital angular momentum, they will, in general, couple to this via spin-orbit coupling (SO). As this effect is rarely incorporated into a Heisenberg model we will now introduce the corresponding operator.

\subsection{Spin-Orbit Coupling}
Spin-orbit (SO) coupling has previously been used in the $UV$-Hubbard model \cite{Barford:2010aa} and we can map the Hamiltonian onto the Heisenberg model (see App. A) to obtain,
\begin{equation}
\hat{H}_{\text{SO}} = A\sum_{i=1}^{N-1} \left(\sum_{\sigma = \uparrow,\downarrow} \hat{n}_{i,\sigma}\hat{n}_{i+1,\sigma} - \frac{1}{\hbar^2}\left(\hat{S}_i^+\hat{S}_{i+1}^+ + \hat{S}_i^-\hat{S}_{i+1}^-\right)\right),
\end{equation}
where $\hat{n}_{i\sigma}$ is the number operator for spins in state $\sigma$ on site $i$ and $\hat{S}_i^{\pm} = \spin{x}{i}\pm\ig\spin{y}{i}$ are spin-flip operators for site $i$. The SO Hamiltonian couples explicitly states which differ in exactly one pair of adjacent spins that are in the same spin state. For instance, it couples the $\ket{\uparrow\uparrow}$ and $\ket{\downarrow\downarrow}$ states of a two-spin system. For the minimal size problem of four spins it couples the singlet with the quintet states, while the triplets only couple to each other. For instance the state $\ket{\uparrow\uparrow\downarrow\downarrow}$ couples to the state $\ket{\uparrow\uparrow\uparrow\downarrow}$ (and others) while state $\ket{\uparrow\downarrow\uparrow\uparrow}$ only couples directly to the state $\ket{\uparrow\downarrow\downarrow\downarrow}$. The singlet-quintet coupling is important for singlet fission, as it can convert the $\TT$ state into a $^5(\text{T}\cdots\text{T})$ state that can then spin-allowed decohere into two uncoupled triplets which can be harvested.

However, spin-orbit coupling is small in most organic semiconductors due to the absence of heavy elements in the molecular structure. In order to make the generation of triplets efficient other effects have to be harnessed. Interactions with an environment, for instance nuclear degrees of freedom or coupling to nuclear and external spins and magnetic fields, is expected to make the generation of triplets more efficient due to a weakening of the spin-conservation requirement. This is discussed in Section III.

\subsection{Basis Set}
\begin{figure}
\includegraphics[width=0.75\linewidth]{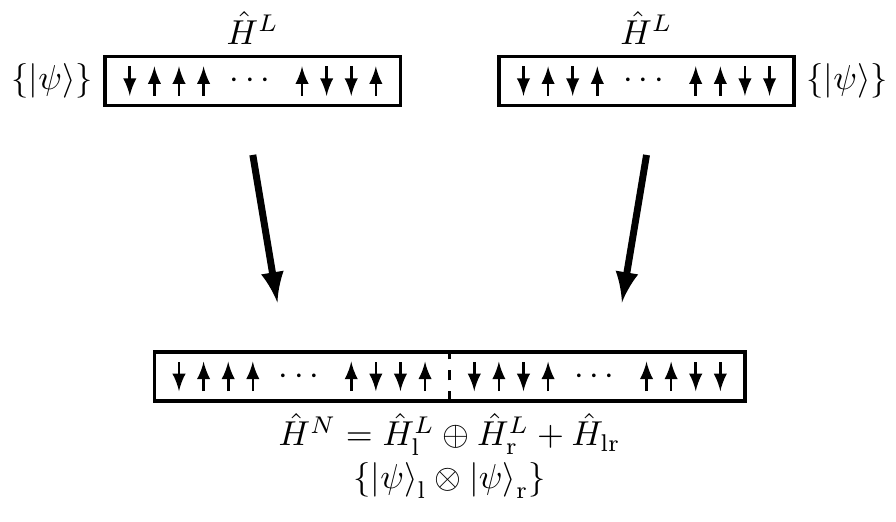}
\caption{\label{Fig:4} The basis set of the full chain of length $N$ is made up of the eigenbasis of the Hamiltonian $\hat{H}^L$ describing the a half-chain. Due to symmetry the eigenbasis sets will be identical for each half-chain and will be indexed with l and r to denote the respective half-chains. If $\hat{H}_{\text{lr}} = 0$ then the direct product of bases is also an eigenbasis of the full chain of length $N$.}
\end{figure}

As discussed above, the $\ket{\TT}$ spanning the entire chain has to decay into two independence triplets for singlet fission to be complete.  As we have a uniform model and no disorder in our system such triplet states will, by necessity, localise on a half-chain and a suitable basis has to be adapted. To this end, we construct the basis of the full chain of length $N$ from the eigenbasis of a Hamiltonian covering the half-chains of length $L=N/2$. Due to the symmetry of our system, this diagonalisation of the half-chain has to be completed only once. Fig. \ref{Fig:4} shows a schematic of how the basis is constructed.
Since $\hat{H}_{\text{lr}} = J\hat{\bf S}_L\cdot\hat{\bf S}_{L+1} \neq 0$ the direct product basis of the half-chains will not be an eigenbasis of the full chain and we find the eigenstates by diagonalising the full Hamiltonian as,
\begin{equation}
\ket{\Psi_n} = \sum_{a,b = 1}^{2^L} \Psi^n_{a,b}\ket{\psi_a}_{\text{l}}\otimes\ket{\psi_b}_{\text{r}}.
\end{equation}
In turn, each half-chain state can be written in a (physical) spin basis as,
\begin{equation}
\ket{\psi_a} = \sum_{j=1}^{2^L} \psi^a_j \ket{{\bm \sigma}_j},
\end{equation}
where $\ket{{\bm \sigma}_j} = \bigotimes_{i=1}^L \ket{\sigma_i}$ is a spin-state in the local basis $\left\{\ket{\sigma_i}\right\}$, which can take values $\left\{\ket{\uparrow},\ket{\downarrow}\right\}$ for each site. These half-chain states can then be used to construct full-chain basis states, such as $\ket{\text{T}_{+1}}_{\text{l}}\otimes\ket{\text{T}_{+1}}_{\text{r}}$, in which we can express the eigenstates of $\hat{H}^N$. Naturally, as these are also eigenstates of $\spin{z}{}$ and $\hat{S}^2$, we can assign spin and magnetic quantum numbers to the full chain states and decompose these within the combined half-chain bases. 

The choice of this basis allows an easy identification of half-chain states combining to an overall state. For instance, the state $\ket{\text{T}_{+1}}_{\text{l}}\otimes\ket{\text{T}_{-1}}_{\text{r}}$ can combine with other states to give a representation of the $\ket{\TT}$ state, highlighting the local triplet character for each half-chain.

The basis set described allows for a natural bipartitioning of the full chain, and the notion of coherence and entanglement can then be readily investigated in this half-chain basis. 

\subsection{Interaction with an Environment}
The quantum-mechanically exact evolution of a system coupled to an environment and described by a density operator $\hat{\rho}_{\text{SE}}$ is given by the Liouville-von Neumann equation, which is the density operator equivalent of the Schr\"{o}dinger equation, and given as,
\begin{equation}
\frac{\text{d}\hat{\rho}_{\text{SE}}}{\text{d}t} \equiv \dot{\rho}_{\text{SE}} = -\frac{\ig}{\hbar}\left[\hat{H}_{\text{SE}},\hat{\rho}_{\text{SE}}\right],
\label{Eq:LvN}
\end{equation}
where $\hat{H}$ is the Hamiltonian describing the system, bath, and all interactions between them and $[A,B]$ denotes the commutator. By assuming that at time $t=0$ no correlations exist between the system and the bath, and that all subsequent correlations are short-lived (the Born and Markov approximations, respectively) an effective equation-of-motion can be derived in which the environment has been traced out, \viz
\begin{equation}
\hat{\rho}_{\text{S}} = \tr_{\text{E}}\hat{\rho}_{\text{SE}} = \sum_{s,s'\in\text{S}} \ket{s}\left(\sum_{e\in\text{E}}\bra{s,e}\hat{\rho}_{\text{SE}}\ket{s',e}\right)\bra{s'},
\end{equation}
where $\left\{\ket{s}\right\}$ and $\left\{\ket{e}\right\}$ denote the basis for the system and the environment, respectively. By transforming the equations into the Interaction Picture the system-environment interaction will give rise to an additional term for Eq. (\ref{Eq:LvN}) \cite{Alicki:2007aa}. The resulting Lindblad master equation has as its first term the Liouville-von Neumann equation for the system degrees of freedom only, while the second term includes the interaction of the system with the environment,
\begin{equation}
\frac{\text{d}\hat{\rho_{\text{S}}}}{\text{d}t} \equiv \dot{\rho}_{\text{S}} = -\frac{\ig}{\hbar}\left[\hat{H}_{\text{S}},\hat{\rho}_{\text{S}}\right] - \frac{\gamma}{2}\sum_m \left(\left\{\hat{L}_m^{\dagger}\hat{L}_m,\hat{\rho}_{\text{S}}\right\} - 2 \hat{L}_m\hat{\rho}_{\text{S}}\hat{L}_m^{\dagger}\right).
\label{Eq:Lindblad}
\end{equation}
Here, $\gamma$ is the Lindblad coefficient, quantifying the strength of the interaction between system and bath, $\hat{L}_m$ is the Lindblad operator which contains the action of the bath on the system, and $\left\{A,B\right\}$ denotes the anti-commutator. Note that the first term only contains information about the system degrees of freedom and as such $\hat{H}_{\text{S}}$ acts on the system alone. We will drop the index S from now to improve legibility and because all environmental degrees of freedom are now implicitly incorporated in the Lindblad operators.

We can use the Lindblad equation to derive equations-of-motion for observables. By noting that,
\begin{equation}
\frac{\text{d}\left\langle O\right\rangle}{\text{d}t} \equiv \dot{\left\langle O\right\rangle} = \tr \left(\dot{\rho}\hat{O}\right),
\label{Eq:Expec}
\end{equation}
and using Eq. (\ref{Eq:Lindblad}) we find,
\begin{equation}
\dot{\left\langle O\right\rangle} = \frac{\ig}{\hbar}\left\langle\left[\hat{H},\hat{O}\right]\right\rangle - \frac{\gamma}{2}\sum_m\left\langle \hat{L}_m^{\dagger}\left[\hat{L}_m,\hat{O}\right]+\left[\hat{O},\hat{L}_m^{\dagger}\right]\hat{L}_m\right\rangle,
\label{Eq:EOM}
\end{equation}
where $\left\langle...\right\rangle$ denotes the expectation value at time $t$ with respect to the state $\rho(t)$. Similarly, if we can find a closed expression for $\dot{\rho}$ then we can use Eq. (\ref{Eq:Expec}) directly to calculate how the observable changes in time, as we will do in Section III.

\subsection{Coherence and Entanglement Measures}
In recent years the study of entanglement and coherence as a resource has drawn considerable attention. Central to these \emph{quantum resource theories} (QRTs)\cite{Baumgratz:2014aa,Chitambar:2019aa} is the notion that a quantum phenomenon is a resource that can be used to create certain states of systems, similar to the more conventional idea of a resource. Within this framework, which goes far beyond entanglement and coherence, states are classified according to whether they are \emph{free} or \emph{prohibited} given a certain operation. QRTs state, for instance, that any decoherent state can be converted into any other decoherent state given what is called \emph{incoherent operations}. But in order to create a coherent state from an incoherent one, the coherence has to be obtained either by creating it given a certain operation, which is not incoherent, or it has to be coupled to a second system or environment which has coherence.

Similar considerations hold for entanglement. Free operations will conserve or destroy entanglement, but never create it. The coupling to an environment or second system once again creates the opportunity to transfer entanglement.

The importance of QRTs, beyond their fundamental interest, is that recasting entanglement and coherence as a resource axiomatically allows these to be measured and quantified. As such we can investigate the \emph{amount} of coherence and entanglement over time. The resulting \emph{quantum measures} on these resources are metrics which decay monotonically with the resource and vanish if none is present in the system. Of the measures proposed for coherence we have chosen the $l_1$-measure \cite{Baumgratz:2014aa} due to its straightforward definition, which allows for calculation by hand,
\begin{equation}
C_{l_1}(t;\hat{\rho}) = \sum_{i\neq j} |\rho_{ij}(t)| = \sum_{i,j} |\rho_{ij}(t)| - 1.
\label{Eq:Cohl1}
\end{equation}
Evidently, this measure vanishes if the off-diagonal entries in the corresponding density matrices vanish, \ie\ if no coherence is present.

Entanglement measures are more difficult to compute, generally, because they do not show in a density matrix as obviously as coherences. The \emph{negativity} of a density matrix which has been partitioned into two sub-systems, A and B, between which we want to measure the entanglement, has been proposed as a numerically accessible entanglement measure\cite{Chitambar:2019aa}, making use of the Horodecki criterion\cite{Horodecki:2009aa}. Its value is given as,
\begin{equation}
N(t;\hat{\rho}_{\text{AB}}) = \left|\left|\left(\hat{\hat{\openone}}_{\text{A}}\otimes \hat{\hat{\mathcal{T}}}_{\text{B}}\right)\hat{\rho}_{\text{AB}}(t)\right|\right|_{\text{tr}} - 1,
\end{equation}
where $\hat{\hat{\openone}}_{\text{A}}$ is the identity operator on subsystem A, $\hat{\hat{\mathcal{T}}}_{\text{B}}$ is the transposition superoperator on subsystem B, and $\left|\left|\cdot\right|\right|_{\text{tr}}$ is the trace norm. For an unentangled system the transposition of sub-system B will not affect the singular values of the full density operator, $\hat{\rho}_{\text{AB}}$, which add up to 1, causing the negativity to vanish. This is no longer true for an entangled system, whose density matrix spectrum will change under partial transposition, giving the negativity a non-zero value, and therefore indicating entanglement between the two sub-systems. Instead of using the negativity directly, we use a logarithmic variant, which is a valid measure because the logarithm is monotonic and we define,
\begin{equation}
E_N(t; \hat{\rho}_{\text{AB}}) = \log_2 \left|\left|\left(\hat{\hat{\openone}}_{\text{A}}\otimes\hat{\hat{\mathcal{T}}}_{\text{B}}\right)\hat{\rho}_{\text{AB}}(t)\right|\right|_{\text{tr}}.
\label{Eq:EntLog}
\end{equation}
For a bipartite system with two states per subsystem this measure is 1 for a maximally entangled state, the logarithm therefore normalises the negativity.

If both coherence and entanglement vanish in time, and we can exclude the possibility of these being generated by an interaction with the environment (see App. B), then we can define the timescale on which singlet fission is truly complete, as the entanglement and coherence between the two triplets vanishes and non-geminate triplets are produced.

\section{Equations of Motion}

The two types of interaction considered here correspond to the Lindblad operators $\hat{L}_m = \spin{\pm}{m}/\hbar$, which is a local spin-flip environment, and $\hat{L}_m = \spin{z}{m}/\hbar$, which induces spin dephasing in the system. We will now discuss both cases briefly before presenting numerical results. 

\subsection{Longitudinal Relaxation}
One choice of environmental interaction is with (local) magnetic fields inducing spin flips. In this case $\hat{L}_m = \spin{+}{m}/\hbar$ or $\hat{L}_m = \spin{-}{m}/\hbar$ or a statistical mixture of both, for instance
\begin{equation}
\hat{L}_m = \frac{1}{2\hbar}\left(\spin{+}{m} + \spin{-}{m}\right) = \frac{1}{\hbar}\spin{x}{m},
\label{Eq:effL}
\end{equation}
for equal probabilities. It is shown in App. B that this operation is an incoherent operation.

Using Eq. (\ref{Eq:EOM}) we can find the equations-of-motion for observables. For instance, the local magnetisation on site $i$ evolves as,
\begin{widetext}
\begin{equation}
\dot{\left\langle S^{z}_{i}\right\rangle}_{\parallel} = -\frac{J}{\hbar}\left\langle\left[{\bf S}_i\times{\bf S}_{i-1} + {\bf S}_i \times{\bf S}_{i+1}\right]^z\right\rangle - \ig \frac{A}{\hbar} \left\langle  S_i^+S_{i-1}^+ -  S_i^-S_{i-1}^- +  S_i^+S_{i+1}^+ - S_i^-S_{i+1}^-\right\rangle-\gamma\left\langle S^z_i\right\rangle,
\end{equation}
\end{widetext}
where $\parallel$ denotes the case for longitudinal relaxation. Here, the first term is the precession of the spin around the quantisation axis, the second term is due to the spin-orbit coupling, flipping parallel spins on neighbouring sites, and the last term is longitudinal relaxation (indicated by $\parallel$), which is equivalent to that observed in magnetic resonance experiments with a relaxation time of $T_1 = \gamma^{-1}$. We can find the total magnetisation of the sample by summing over the sites,
 \begin{widetext}
\begin{equation}
\dot{\left\langle S^z\right\rangle}_{\parallel} = -2\frac{J}{\hbar}\sum_{i=1}^{N-1}\left\langle \left[{\bf S}_i\times{\bf S}_{i+1}\right]^z\right\rangle-2\ig \frac{A}{\hbar}\sum_{i=1}^{N-1}\left\langle \spin{+}{i}\spin{+}{i+1}-\spin{-}{i}\spin{-}{i+1}\right\rangle - \gamma\left\langle S^z\right\rangle.
\end{equation}
\end{widetext}
Unsurprisingly, the effect of the spin-flipping environment is that the magnetisation of the system is not conserved. We will come back to this in Section IV, when discussing numerical results and steady states.

The effect of the spin-flip interaction of the system with an environment is closely related to the effect of spin-orbit coupling, as both allow for spins to be flipped, leading to a change in the magnetisation of the sample. As the spin-orbit coupling is part of the system Hamiltonian, and therefore incorporated within the model if $A\neq 0,$ we can use a different system-environment interaction to model a more complex (and realistic) situation, namely transverse relaxation.

\subsection{Transverse Relaxation}
\subsubsection{Spin Observables}
A different choice of environmental interaction is to consider other spins coupled to each site, for instance nuclear spins or other electrons, by using $\hat{L}_m = \spin{z}{m}/\hbar$. As above, it is straightforward to show that this is a free operation as shown in App. B. The action of $\hat{L}_m = \spin{z}{m}/\hbar$ is to give any spin in state $\ket{\downarrow}$ a phase of $\pi$ while leaving spins in state $\ket{\uparrow}$ unchanged. Two neighbouring $\ket{\uparrow\downarrow}$ spins dephase accordingly. The resulting mixed state will show less coherence between states in the half-chain basis, as we will discuss below. The dephasing can also be understood in terms of the expectation values of the spin-operators. The magnetisation is constant under this environment, while the $S_x$ and $S_y$ values decay in time. We can once again investigate the magnetisation of the system as it evolves in time by deriving the equation of motion. We find,
\begin{equation}
\dot{\left\langle S^z\right\rangle}_{\perp} = -2\frac{J}{\hbar}\sum_{i=1}^{N-1}\left\langle \left[{\bf S}_i\times{\bf S}_{i+1}\right]^z\right\rangle-2\ig \frac{A}{\hbar}\sum_{i=1}^{N-1}\left\langle \spin{+}{i}\spin{+}{i+1}-\spin{-}{i}\spin{-}{i+1}\right\rangle,
\end{equation}
which is identical to the longitudinal relaxation, but without damping the term. Here, $\perp$ denotes the transverse relaxation case. The effect of this absence of damping is the restriction of possible $S^z$-values if $A=0$. However, even if $A\neq 0$ the possible values of $S_z$ are still partially restricted, as spin-orbit coupling will flip pairs of parallel spins, changing the magnetisation of the system in steps of $\pm \hbar$.

For the other components of the spin-vector we can find, for instance,
\begin{widetext}
\begin{equation}
\dot{\left\langle S^x\right\rangle}_{\perp} = -2\frac{J}{\hbar}\sum_{i=1}^{N-1}\left\langle \left[{\bf S}_i\times{\bf S}_{i+1}\right]^x\right\rangle-2\ig \frac{A}{\hbar}\sum_{i=1}^{N-1}\left\langle \spin{y}{i-1}\spin{z}{i}-\spin{z}{i}\spin{y}{i+1}\right\rangle - \frac{\gamma}{2}\left\langle S^x\right\rangle,
\label{Eq:23}
\end{equation}
\end{widetext}
showing a damping term for the transverse components, akin to transverse relaxation in magnetic resonance experiments with a decay time of $T_2 = 2\gamma^{-1}$, and indicated by $\perp$.

\subsubsection{Steady States}
Before describing our numerical results, we now discuss the steady state solutions of the Lindblad equation, Eq. (\ref{Eq:Lindblad}). This has several stationary states which correspond to the kernel of the augmented Liouvillian, \ie,
\begin{widetext}
\begin{equation}
\dot{\rho} = -\frac{\ig}{\hbar}\left[\hat{H},\hat{\rho}\right] - \frac{\gamma}{2}\sum_{m=1}^N \left(\left\{\hat{L}_m^{\dagger}\hat{L},\hat{\rho}\right\} - 2\hat{L}_m\hat{\rho}\hat{L}_m^{\dagger}\right) = \hat{\hat{\mathcal{L}}}\hat{\rho} + \hat{\hat{\mathcal{D}}}\hat{\rho} = \left(\hat{\hat{\mathcal{L}}}+\hat{\hat{\mathcal{D}}}\right)\hat{\rho},
\label{Eq:kernel1}
\end{equation}
\end{widetext}
where $\hat{\hat{\ }}$\ indicates a superoperator. Vectorising the density matrix then shows the similarity with the kernel when finding the stationary states as we require the solution to,
\begin{equation}
\dot{\vec{\rho}} = \left(\hat{\mathcal{L}} + \hat{\mathcal{D}}\right)\vec{\rho} = \vec{0},
\label{Eq:kernel2}
\end{equation}
where $\vec{\ }$ indicates a vectorisation. It is evident from Eqs. (\ref{Eq:kernel1}) and (\ref{Eq:kernel2}) that the condition for a stationary state is that the effects of the unitary evolution and the dissipator have to cancel out, or both have to vanish. It is easier to find conditions under which both contributions vanish. The unitary contribution, $\hat{\hat{\mathcal{L}}}\hat{\rho}$, vanishes if $\hat{\rho}$ is in some linear combination of the eigenstates of $\hat{H}$. Let $\ket{\sigma}\bra{\sigma}$ be such eigenstates, then any states which can be written as,
\begin{equation}
\hat{\rho} = \sum_{\sigma=1}^N \rho_{\sigma}\ket{\sigma}\bra{\sigma},
\end{equation}
where $N$ is the dimensionality of the Hilbert space associated with $\hat{H}$, will commute with $\hat{H}$ and $\hat{\hat{\mathcal{L}}}\hat{\rho}_0 =0$. 

The dissipator, $\hat{\hat{\mathcal{D}}}\hat{\rho}$, for such a state yields,
\begin{widetext}
\begin{equation}
\hat{\hat{\mathcal{D}}}\hat{\rho} = -\frac{\gamma}{2}\sum_{\sigma=1}^N\rho_{\sigma}\sum_m \left(\hat{L}_m^{\dagger}\hat{L}_m\ket{\sigma}\bra{\sigma}+\ket{\sigma}\bra{\sigma}\hat{L}_m^{\dagger}\hat{L}_m - 2\hat{L}_m\ket{\sigma}\bra{\sigma}\hat{L}_m^{\dagger}\right).
\end{equation}
\end{widetext}
This sum can, naturally, vanish in many different ways and it depends on the nature of $\hat{L}_m$ and the states $\ket{\sigma}\bra{\sigma}$ if it does or does not. However, we can easily draw up some conditions under which the sum will vanish. We could impose a stronger condition, namely, that each term vanishes, and as such,
\begin{equation}
\hat{L}_m^{\dagger}\hat{L}_m\ket{\sigma}\bra{\sigma} + \ket{\sigma}\bra{\sigma}\hat{L}_m^{\dagger}\hat{L}_m = 2\hat{L}_m\ket{\sigma}\bra{\sigma}\hat{L}_m^{\dagger},
\end{equation}
which is fulfilled if $\hat{L}_m = \hat{L}_m^{\dagger}$, \ie\ if $\hat{L}_m$ is hermitian, and if $\left[\ket{\sigma}\bra{\sigma},\hat{L}_m\right] = 0$, \ie\ if $\ket{\sigma}\bra{\sigma}$ is also an eigenstate of $\hat{L}_m$. Thus, if $\hat{L}_m$ is hermitian and shares an eigenset with $\hat{H}$, then any diagonal state is a stationary state of the Lindblad equation.  We can show this explicitly by decomposing the density matrix into unitary and dissipating contributions, \viz
\begin{equation}
\begin{split}
\dot{\rho} &= \hat{\hat{\mathcal{L}}}\hat{\rho} + \hat{\hat{\mathcal{D}}}\hat{\rho} = \dot{\rho}^{\text{uni}} + \dot{\rho}^{\text{diss}}.
\end{split}
\end{equation}
The second term acts as a damping term and as the unitary term will induce oscillations between eigenstates of the Hamiltonian, the dissipator will need to force the state into a state that commutes with the Hamiltonian. For a hermitian operator $\hat{L}_m$ the dissipation term can be written as,
\begin{equation}
\dot{\rho}^{\text{diss}} = -\frac{\gamma}{2}\sum_m \left(\hat{L}_m^2\hat{\rho} + \hat{\rho}\hat{L}_m^2 - 2\hat{L}_m\hat{\rho}\hat{L}_m\right).
\end{equation}
If $\hat{L}_m$ is involutary (\ie\ $\hat{L}_m^2 =\hat{\openone}$) then the first two terms can be compressed and we find,
\begin{equation}
\dot{\rho}^{\text{diss}} =  -\gamma\sum_m \left(\hat{\rho} - \hat{L}_m\hat{\rho}\hat{L}_m\right).
\end{equation}
$\hat{L}_m = \spin{z}{m}/\hbar$ fulfils the condition set out above (with a prefactor of $1/4$). We can now decompose the density matrix into a diagonal and an off-diagonal part,
\begin{equation}
\hat{\rho} = \hat{\rho}_{\text{diag}} + \hat{\rho}_{\text{off}} = \sum_i \rho_{i}\ket{i}\bra{i} + \sum_{i\neq j}\rho_{ij}\ket{i}\bra{j},
\end{equation}
where $\rho_i$ are the populations and $\sum_i \rho_i = 1$. If we choose the basis states to be eigenstates of the on-site $\spin{z}{m}$ operators, \ie\ we adopt a local spin basis, then we can write,
\begin{widetext}
\begin{equation}
\dot{\rho}^{\text{diss}} = -\frac{\gamma}{4}\sum_m\hat{\rho} + \frac{\gamma}{\hbar^2}\sum_m \spin{z}{m}\left(\sum_i \rho_i \ket{i}\bra{i} + \sum_{i\neq j} \rho_{ij}\ket{i}\bra{j}\right)\spin{z}{m}.
\end{equation}
\end{widetext}
Due to the hermiticity of $\spin{z}{m}$ the operator acts on the bra and ket in equal ways and will produce an eigenvalue of $\pm\hbar^2/4$. We then find,
\begin{widetext}
\begin{equation}
\dot{\rho}^{\text{diss}} = -\frac{\gamma}{4}\sum_m\hat{\rho} + \frac{\gamma}{\hbar^2}\sum_m \left(\sum_i \rho_i \left(S_{m,i}^z\right)^2\ket{i}\bra{i} + \sum_{i\neq j} \rho_{ij}\spin{z}{m}\ket{i}\bra{j}\spin{z}{m}\right).
\end{equation}
\end{widetext}
But, $\left(S_{m,i}^z\right)^2 = \frac{\hbar^2}{4}$ for any site in any state and we can re-write,
\begin{equation}
\dot{\rho}^{\text{diss}} = -\frac{\gamma}{4}\sum_m\hat{\rho} + \frac{\gamma}{4}\sum_m \hat{\rho}_{\text{diag}} + \frac{\gamma}{\hbar^2}\sum_m\sum_{i\neq j} \rho_{ij}\spin{z}{m}\ket{i}\bra{j}\spin{z}{m},
\end{equation}
or
\begin{equation}
\begin{split}
\dot{\rho}^{\text{diss}} &= -\frac{\gamma}{4}\sum_m\left(\hat{\rho}-\hat{\rho}_{\text{diag}}\right) + \frac{\gamma}{\hbar^2}\sum_m \sum_{i\neq j} S_{m,i}^zS_{m,j}^z\rho_{ij}\ket{i}\bra{j},\\
&= -\frac{\gamma N}{4}\hat{\rho}_{\text{off}} + \gamma\sum_{i\neq j}S_{ij}\rho_{ij}\ket{i}\bra{j},
\end{split}
\end{equation}
where $S_{ij} = \frac{1}{\hbar^2}\sum_m S_{m,i}^zS_{m,j}^z$.
The last term can be simplified by using the Hadamard product, which corresponds to element-wise multiplication of matrices, denoted $\odot$. In matrix representation the last line then takes the form,
\begin{equation}
\dot{\bm \rho}^{\text{diss}} = -\frac{\gamma N}{4}{\bm \rho}_{\text{off}} + \gamma {\bm S}\odot{\bm \rho}_{\text{off}}.
\label{Eq:EOM2}
\end{equation}
The form of Eq. (\ref{Eq:EOM2}) shows that the dissipator will act on each off-diagonal element individually (\ie\ no convolution of elements) and not on the diagonal elements at all. By projecting out a given element this becomes more obvious, \viz,
\begin{widetext}
\begin{equation}
\dot{\rho}_{ij}^{\text{diss}} = \bra{i}\dot{\bm\rho}^{\text{diss}}\ket{j} = -\frac{\gamma N}{4}\rho_{ij} +\gamma S_{ij}\rho_{ij} = \left(\gamma S_{ij} - \frac{\gamma N}{4}\right)\rho_{ij}.
\label{Eq:C15}
\end{equation}
\end{widetext}
The value of $S_{ij}$ is determined by the number of aligned spins between the two states $\ket{i}$ and $\ket{j}$. In the case of perfect alignment, which corresponds to a diagonal entry ($i=j$) in the density matrix, we have $S_{ii} = N/4$ and the decay rate is 0, showing that diagonal entries are not affected by the dissipator. For the case of perfect anti-alignment (which is the anti-diagonal of the density matrix and we will denote such a state by $\ket{i}\bra{\bar{i}}$) we have $S_{i\bar{i}} = -N/4$, showing that anti-diagonal entries in the density matrix vanish the most quickly. All other states will have decay rates between 0 and $-N/2$ varying in steps of $1/2$. The more aligned states are, \ie\ the fewer spin flips have to be performed to convert state $\ket{i}$ into $\ket{j}$, the longer-lived any coherence between the states will be.

This can be illustrated for the $N=4$ case: for instance coherence between the states $\ket{\text{T}_{+1}}_{\text{l}}\ket{\text{T}_{+1}}_{\text{r}}$ and $\ket{\text{T}_{-1}}_{\text{l}}\ket{\text{T}_{-1}}_{\text{r}}$ will decay with a rate of $-2\gamma$, which is the maximum decay rate. 

\subsubsection{Quantum Measures}

We can solve the differential equations in Eq (\ref{Eq:C15}) straightforwardly to find,
\begin{equation}
\rho_{ij}^{\text{diss}}(t) = \rho_{ij}^{(0)} \ex^{-\Gamma_{ij} t},
\label{Eq:decay}
\end{equation}
where $\Gamma_{ij}$ is the entry-dependent decay rate which is 0 for $i=j$ and otherwise the prefactor in Eq. (E10).

We can use this result to estimate the decay of observables and quantum measures. The $l_1$-measure of coherence (see Eq. (\ref{Eq:Cohl1})) is then given by,
\begin{equation}
C_{l_1}(t;\hat{\rho}) = \sum_{i\neq j}\rho_{ij}(t) = \sum_{i\neq j} \rho_{ij}(0)\ex^{-\Gamma_{ij} t},
\end{equation}
The slowest decay rate will naturally bound the value of $C_{l_1}$ from above and this decay rate is applicable to coherence elements that share only one aligned spin between the two states and is $\gamma/2$ and hence,
\begin{equation}
C_{l_1}(t;\hat{\rho}) \ = \sum_{i \neq j}\rho_{ij}(0)\ex^{-\Gamma_{ij} t} \leq C_{l_1}(0;\hat{\rho})\ex^{-\frac{\gamma t}{2}}.
\label{Eq:upper}
\end{equation}

We can therefore see that $\gamma$ influences both observables and the coherence measure in the same way, making the former a potential proxy for quantum effects.

We can also show that the entanglement measure presented in Eq. (\ref{Eq:EntLog}) will decay exponentially if the off-diagonal elements do so. Consider an entangled state of a two-spin system, \ie\ a generalised Bell state, which can be written as,
\begin{equation}
\bm{\rho} = \begin{pmatrix} 0 & 0 & 0 & 0 \\ 0 & \rho_{22} & \rho_{23} & 0 \\ 0 & \rho_{32} & \rho_{33} & 0 \\ 0 & 0 & 0 & 0 \end{pmatrix},
\end{equation}
where the elements $\rho_{ij}$ are now functions of time. We need to calculate the trace-norm of the partially transposed system, which corresponds to the sum of the singular values. We find the negativity then as,
\begin{equation}
E_N(t;\bm{\rho}) = \log_2\left(\rho_{22}(t) + \rho_{33}(t) + 2\sqrt{\rho_{23}(t)\rho_{32}(t)}\right) ,
\end{equation}
and as the state is pure before transposition and will remain so throughout the interaction we have $\rho_{22}(t) + \rho_{33}(t) = 1$ and hence,
\begin{equation}
E_N(t;\bm{\rho}) = \log_2\left(1 + 2\sqrt{\rho_{23}(t)\rho_{32}(t)}\right).
\end{equation}
If the off-diagonal elements decay exponentially with some decay rate $\Gamma$ then,
\begin{equation}
E_N(t;\bm{\rho}) = \log_2\left(1 + 2\sqrt{\rho_{23}(0)\rho_{32}(0)}\ex^{-\Gamma t}\right),
\end{equation}
showing that the entanglement also decays exponentially. We can once again take the slowest decay rate to find an upper bound and hence,
\begin{equation}
E_N(t;\bm{\rho}) \leq \log_2\left(1 + 2\sqrt{\rho_{23}(0)\rho_{32}(0)}\ex^{-\frac{\gamma}{2} t}\right),
\end{equation}

These environmental effects on the systems described above have been numerically simulated and we present the results in the following section.

\section{Numerical Results}
As we are interested not just in the time evolution of observables, but also in quantum effects, a simulation of the entire density matrix of a system under investigation is required. For an $N$ site full-chain the Hilbert space has dimensionality $2^N$ and as such $2^{2N}$ equations of motion (one for each matrix element) have to be integrated. Combined with the construction of our basis, only a limited number of system sizes are accessible via numerical integration of all equations. Using a fourth-order Runge-Kutta method we were able to solve the minimum-size problem of $N=4$, as well as $N=8$ and $N=12$. The Lindblad coefficient, $\gamma$, has units of inverse time and all results are presented in re-scaled time, $\tau = t\gamma$, where $t$ is the physical time. The value of $\gamma$ describes the coupling strength between the system and environment and is a non-trivially determined parameter. 

All numerical simulations have been performed with the dephasing environment ($\hat{L}_m = \spin{z}{m}/\hbar$), leading to transverse relaxation. We chose this environment as the effects from longitudinal relaxation are incorporated into the model via spin-orbit coupling.

We will now first discuss the influence of spin-orbit coupling on the observables using the $N=4$ case as an example.

\subsection{Spin-Orbit Coupling}
\begin{figure}
\centering
\includegraphics[width=0.7\linewidth, clip=true, trim = 0 0 30pt 0]{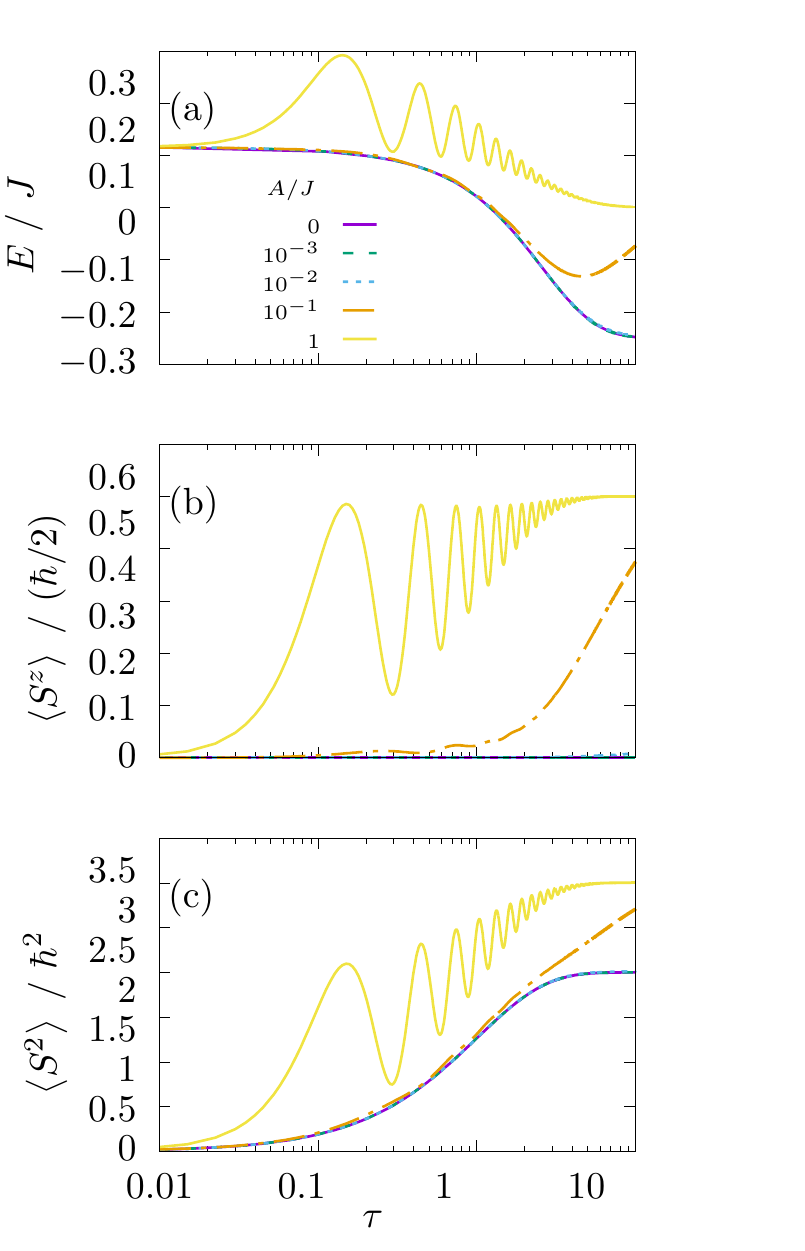}
\caption{\label{Fig:5} Evolution of (a) energy, (b) magnetisation, and (c) spin as a function of reduced time, $\tau = t\gamma$ for the $N=4$ case and varying values of $A/J$.}
\end{figure}

The numerical results for the case of $N=4$ with varying values of $A/J$ are plotted for the total energy, magnetisation, and spin in Fig. \ref{Fig:5}. As we initialise the system in the $\ket{\TT}$ state the energy is initially positive, while both spin and magnetisation have a value of 0. For small values of $A/J \leq 0.01$ the evolution of all three observables is monotonic: the energy decreases to a value of -0.25$J$, the magnetisation remains close to 0, while $\left\langle S^2\right\rangle$ increases to $2\hbar^2$ . Changing $A/J$ to larger values leads to significant changes in the time-evolution of these observables. The energy now approaches 0, while the magnetisation converges on $\hbar / 4$ and the spin approaches values of 3$\hbar^2$. When $A\simeq J$ we also see oscillations for the short-time regime that are damped out over time.

The explanation of these results is as follows: if $A\ll J$ the perturbation on the system is negligible. As we initialise the system in a singlet state, all interactions within the system and with the environment conserve magnetisation. As a consequence, only six states of the $N=4$ Hilbert space are accessible, namely all states with $S^z = 0$. For a chain of four sites there are 16 states: two singlets, three sets of triplets, and one set of quintets. Conserving magnetisation, the initial state can evolve into the singlets, the three triplets with $M_S = 0$, and one quintet with $M_S = 0$. Averaging over the respective values of $S^2$ for these states yields $\left\langle S^2\right\rangle = 2\hbar^2$, which is the limiting value we find for long times. This implies that the state has evolved into a statistical mixture of the states described. Indeed, the diagonal density matrix elements for $\tau  > 10$ are 1/6 for those elements, while 0 for all others. Tab. \ref{Tab:N4States} shows all eigenstates of the $N=4$ chain with their respective observables and energy. 

\begin{table}
\centering
\begin{tabular}{c|c|cc|c}
&$n$ & $\left\langle S^2\right\rangle / \hbar^2$ & $\left\langle S^z\right\rangle / (\hbar/2)$ & $\left\langle E\right\rangle / J$ \\\hline
 S$_1$ & 1 & 0 & 0 & -1.61\\
T$_1$ & 2-4 & 2 & -1,0,+1 & -0.96 \\
T$_2$ & 3-7 & 2 & -1,0,+1 & -0.25 \\
S$_2$ & 8 & 0 & 0 & 0.12 \\
T$_3$ & 9-11 & 2 & -1,0,+1 & 0.46 \\
Q$_1$ & 12-16 & 6 & -2,-1,0,+1,+2 & 0.75\\
\end{tabular}
\caption{Eigenstates of the uniform Heisenberg chain for $N=4$ with expectation values of the total spin, magnetisation, and energy. The labels denote singlet (S), triplet (T), and quintet (Q) multiplicity. Note that the $\ket{\TT}$ state is the S$_2$ state.}
\label{Tab:N4States}
\end{table}

With increasing influence of the spin-orbit coupling, $S^z$ is no longer conserved and within the physical site basis neighbouring spins in the same state can be flipped, changing $S^z$ in steps of $\pm \hbar$. As a result, the quintet states with $M_S = \pm 2$ become accessible and the expectation values of the observables change accordingly, as shown by numerical results.

\begin{figure}[t]
\centering
\includegraphics[width=0.6\linewidth, clip=true, trim = 0 0 30pt 0]{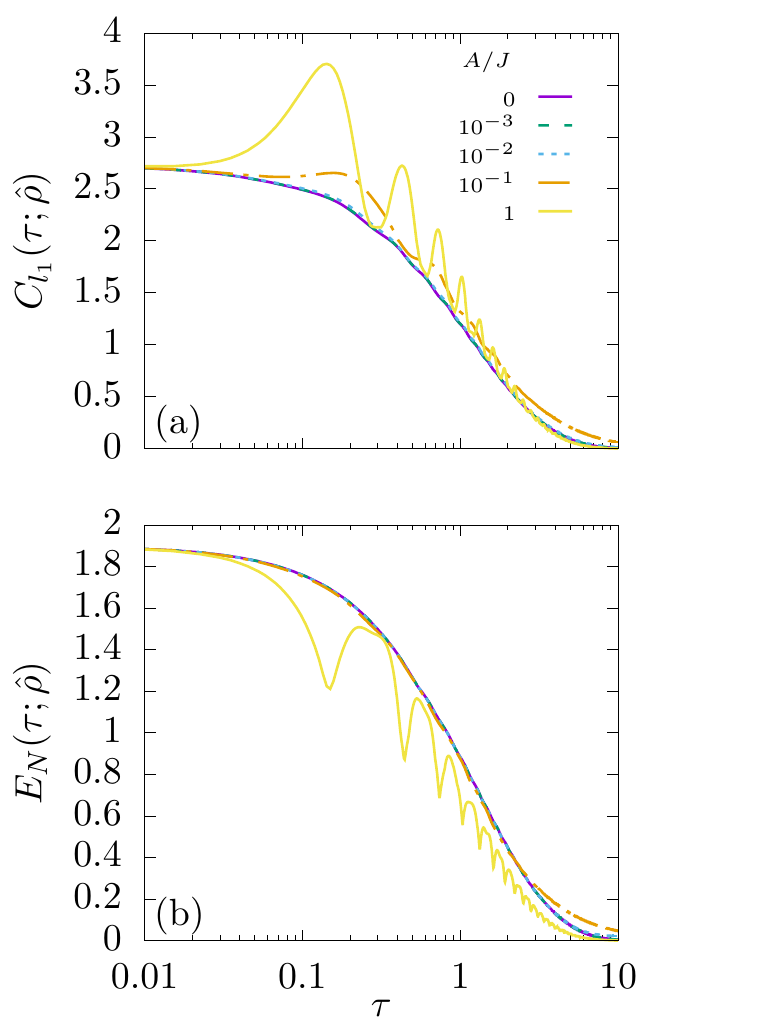}
\caption{\label{Fig:6} Evolution of (a) coherence, and (b) entanglement measures as a function of reduced time, $\tau$, for different values of $A/J$.}
\end{figure}

Fig. \ref{Fig:6} shows the coherence and entanglement measures as defined in Eqs. (\ref{Eq:Cohl1}) and (\ref{Eq:EntLog}) where for the latter the bipartioning into the two half-chains was chosen. Both measures decay rapidly with $\tau$, with the coherence and entanglement having reduced by about 50\%  by $\tau = 1$, and only for $A\simeq J$ can an effect of the spin-orbit coupling be seen, namely in non-monotonic behaviour.

The robustness against the spin-orbit coupling of these effects implies that their decay is primarily caused by coupling to the environment. As the environmental effects also largely affect the change in the observables, a link between the observables and the quantum measures can be established.

\begin{figure}[t]
\centering
\includegraphics[width=0.65\linewidth]{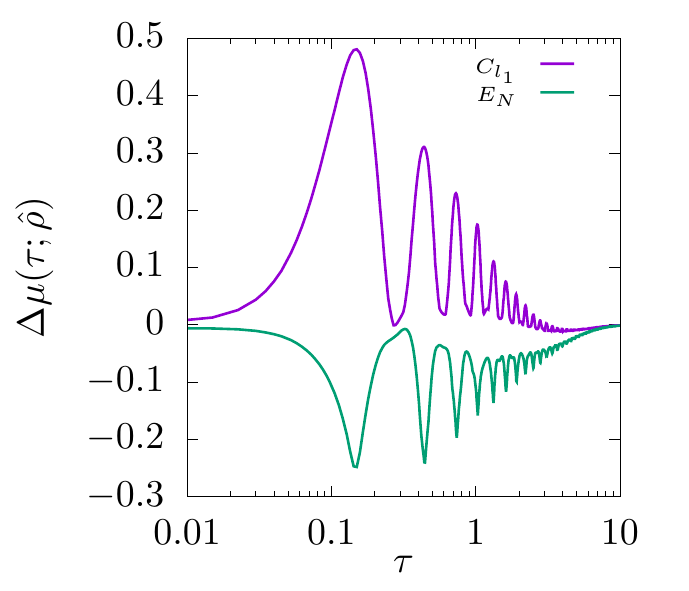}
\caption{\label{Fig:14} Relative, normalised change for the coherence and entanglement measures for $A/J = 1$ and $N=4$. Clearly visible is the synchronised change in both measures, with opposite sign, indicating interconversion.}
\end{figure}

Another feature that can be seen for $A/J = 1$ is the synchronous change in the coherence and entanglement measures. Fig. \ref{Fig:14} shows a normalised version of these measures, defined by
\begin{equation}
\Delta \mu(\tau;\hat{\rho}) = \frac{\mu_1(\tau;\hat{\rho}) - \mu_0(\tau;\hat{\rho})}{\mu_1(0;\hat{\rho})}, 
\end{equation}
where $\mu$ is either of $C_S$ or $E_N$ and the index 1 and 0 denote the cases for $A/J =1$ and $0$. The subtraction of the $A/J = 0$ case ensures the removal of the underlying exponential decay as far as possible. It is evident that coherence increases when entanglement decreases and vice versa. This is in agreement with recently reported theoretical work \cite{Streltsov:2015aa,Zhu:2017ab}, indicating that these two resources can be converted into each other and that spin-orbit coupling can aid this process.

While instructive, $N=4$ is not representative of most systems that undergo singlet fission. In particular, carotenoids are significantly larger systems with $N$ of around 18. While a system that size is beyond our study here, we will now present numerical results for the $N=8$ and $N=12$ cases, choosing both cases of $A=0$ and $A=J$.

\subsection{Size Effects}
With increasing system size the number of available states increases, as does the number of states with $\left\langle S^z\right\rangle = 0,\pm\hbar,...$. The arguments laid out in Section IV.A still hold, but the numbers for the observables have to be adjusted.

\begin{figure}[t]
\centering
\includegraphics[width=0.7\linewidth, clip=true, trim=0 0 30pt 0]{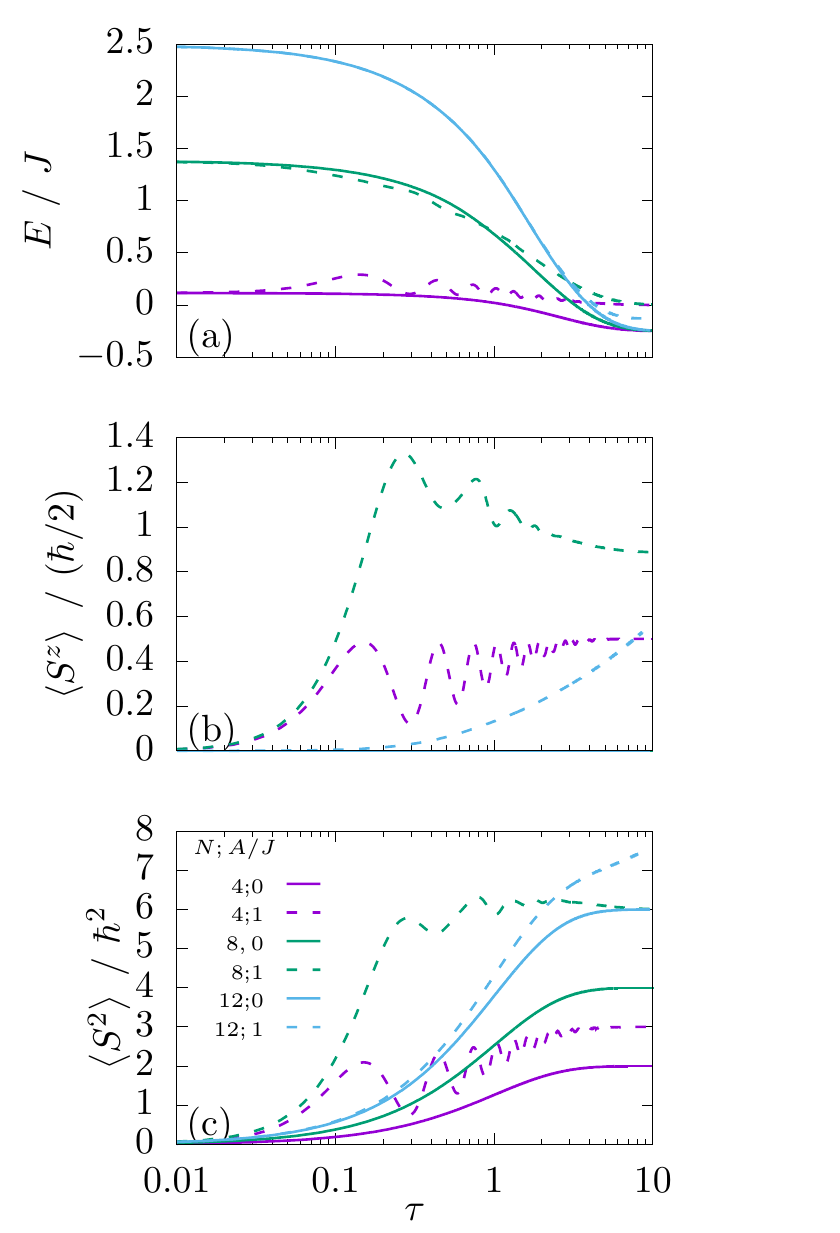}
\caption{\label{Fig:7} Evolution of energy, magnetisation, and spin as a function of reduced time, $\tau$, and $A/J$ for different systems. The colour corresponds to the system size ($N=4$: purple, $N=8$: green, $N=12$: blue) while the line type corresponds to the value of $A/J$, solid for 0 and dashed for 1.}
\end{figure}

Fig. \ref{Fig:7} shows the observables for chains of size 4, 8, and 12 for the two cases of $A/J$.
As can be seen, for increasing $N$ the oscillations for $A/J = 1$ are increasingly damped out. In general, $A/J = 1$ tends to raise the energy of the final state compared to the $A/J = 0$, for which the stationary state reaches $E=-0.25J$, regardless of system size. With an increase in the number of sites the Hilbert space grows exponentially and, as a result, the number of system accessible states grows as well. While for $A/J = 0$ an equal population of these states always yields the same energy; for $A/J = 1$ we see a size dependence, as for increasing $N$ the energy difference of the stationary states for $A/J = 1$ and $A/J = 0$ will decrease. For the other observables we see a similar damping of oscillations. With increasing system size the value of $\gamma$ for critical damping decreases, that is for equal system-environment interaction strength, a larger system will be increasingly over-damped, eliminating the oscillations for short time-scales.

\begin{figure}
\centering
\includegraphics[width=0.6\linewidth, clip=true, trim=0 0 30pt 0]{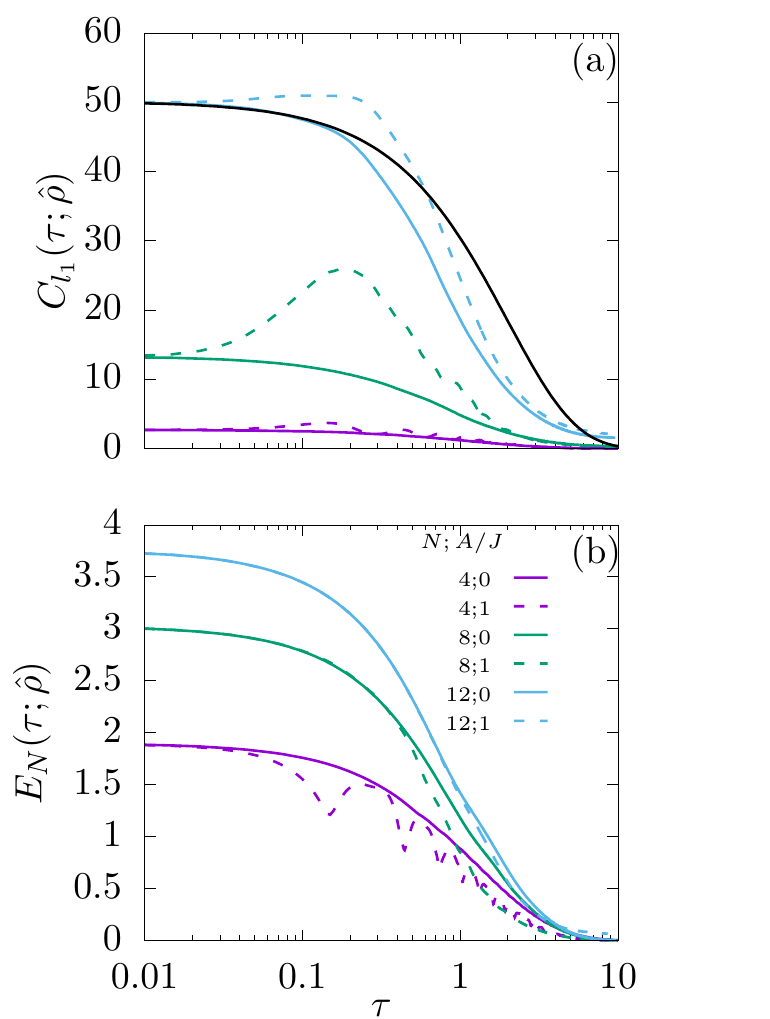}
\caption{\label{Fig:8} Evolution of (a) coherence and (b) entanglement as a function of $\tau$, and for different $A/J$ for different systems. The colour corresponds to the system size ($N=4$: purple, $N=8$: green, $N=12$: blue) while the line type corresponds to the value of $A/J$, solid for 0 and dashed for 1. The residual coherence for long $\tau$ is due to numerical instability of the  integration of the equations of motion. The black line is the upper bound according to Eq. (\ref{Eq:upper}).}
\end{figure}

Examining the coherence and entanglement measures (as shown in Fig. \ref{Fig:8}), we also see the effects of over-damping with increased system size. However, we also observe a secondary size effect: while $N=4$ and $N=8$ are in the under-damped regime, the introduction of spin-orbit coupling has a strong effect on the coherence measure. For $N=12$ this effect is damped out. Spin-orbit coupling therefore enhances coherences in the under-damped regime ($N=4,8$) on short time scales $\tau < 1$. Similarly, increasing system size increases the effect of environmental damping on the coherence. Regardless of size and spin-orbit coupling, all curves show a decay of coherence to about 20\% of their initial value by $\tau = 1$. The black curve indicates the theoretical upper bound for $N=12$ and $A/J=0$ from Eq. (\ref{Eq:upper}). We can see that for both very early and long times the upper bound reproduces the behaviour of the coherence measure, while it decays more rapidly for intermediate times, indicating that off-diagonal terms with decay rates larger than $\gamma/2$ have an increased effect. The discrepancy for $A/J = 1$ for short times is solely due to the spin-orbit coupling. For long $\tau$ this effect is damped out and, in fact, becomes less important for longer chains, in line with a cross-over into the over-damped regime with increasing $N$.

The entanglement, on the other hand, is much more robust to effects of spin-orbit coupling. Only for $\tau>0.5$ does a faster decrease becomes evident with increasing $A$. Spin-orbit coupling therefore seems to increase the coherence on early timescales, but enhances unentanglement of the two half-chains. Again, for long time scales the interaction with the environment becomes dominant, leading to a vanishing of any other effects.

\subsection{Triplet-Triplet Population}
All of the effects observed above indicate a rapid and persistent decoherence and unentanglement of the states on the two half-chains. We can also investigate how the populations of states evolve. In particular, we are interested in the population of those states that have a triplet-triplet character and whether or not they are paired in an overall singlet-state. We can project the full density matrix of the system onto the triplet-triplet subspace by defining the projected density operator
\begin{widetext}
\begin{equation}
\hat{\rho}_{\text{TT}} = \hat{\hat{\mathcal{P}}}_{\text{TT}}\hat{\rho} = \sum_{a,b,c,d} \ket{\text{T}_a}_{\text{l}}\ket{\text{T}_b}_\text{r}\bra{\text{T}_a}_{\text{l}}\bra{\text{T}_b}_{\text{r}}\hat{\rho}\ket{\text{T}_c}_{\text{l}}\ket{\text{T}_d}_{\text{r}}\bra{\text{T}_c}_{\text{l}}\bra{\text{T}_d}_{\text{r}},
\end{equation}
\end{widetext}
where we have suppressed the direct products for clarity. We can define a similar projector for the singlet-singlet (SS) manifold.

\begin{figure}[t]
\centering
\includegraphics[width=0.65\linewidth, clip=true, trim=0 0 30pt 0]{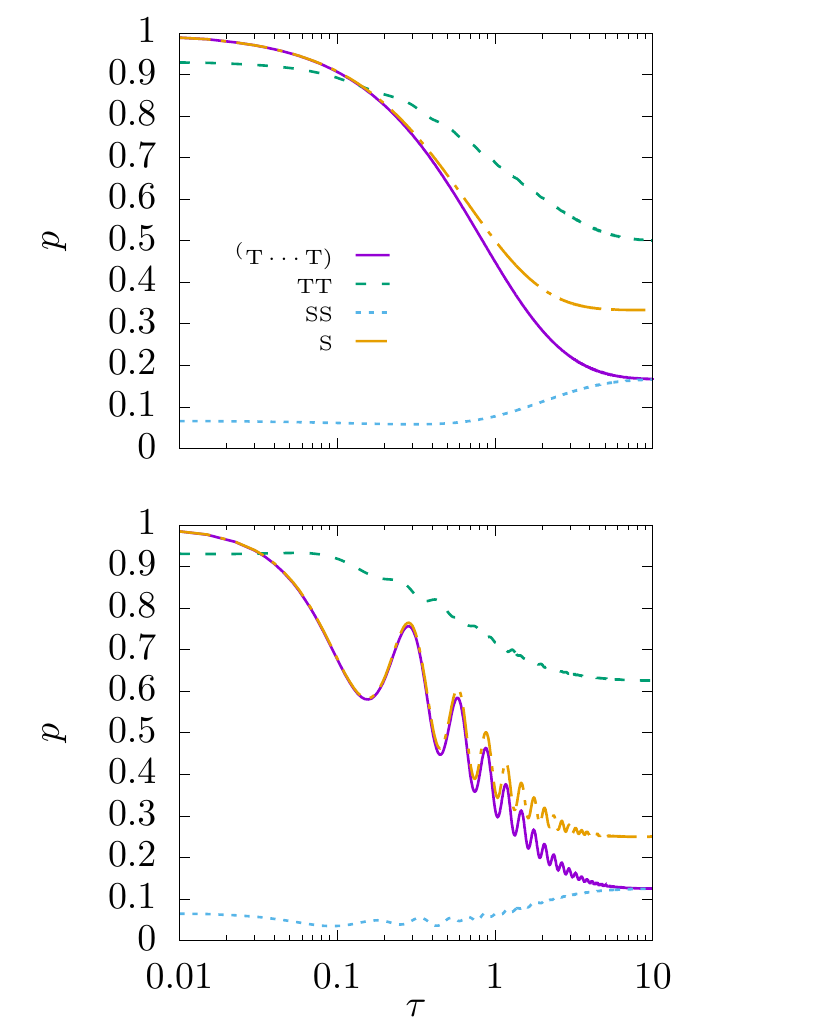}
\caption{\label{Fig:11} Populations of different states and state manifolds for the $N=4$ chain with $A=0$ (top) and $A=J$ (bottom). $\TT$ is the overall-singlet state with local triplet character, TT is the triplet-triplet manifold, S are states of the full chain with singlet character, regardless of the local nature, and SS is the singlet-singlet manifold.}
\end{figure}

Fig. \ref{Fig:11} shows the evolution of different populations for $N=4$ with no and strong spin-orbit coupling. We see that the monotonic decay of the $\TT$ population associated with a decay in the singlet population on the full chain for the $A/J = 0$ case; and oscillating behaviour which is damped out for long $\tau$ for $A/J = 1$. For long times these two curves (\ie S and $\TT$) become distinct, indicating an overall-singlet population without $\TT$ character.

Similarly, we see a decay in the triplet-triplet population, but to a lesser extent than the decay in the $\TT$ population. This indicates a population of higher-spin triplet-triplet states. We see an increase in singlet-singlet states, which becomes identical to the population of the $\TT$ state. This implies that the TT states contributing to the $\TT$ population decay into higher-spin TT states or the SS contribution. Spin-orbit coupling (as shown in Fig. \ref{Fig:11}(b)) increases the population of the TT manifold even further, while also reducing the $\TT$ population, leading to a higher population of high-spin TT states.

\begin{figure}[t]
\centering
\includegraphics[width=0.6\linewidth, clip=true, trim=0 0 30pt 0]{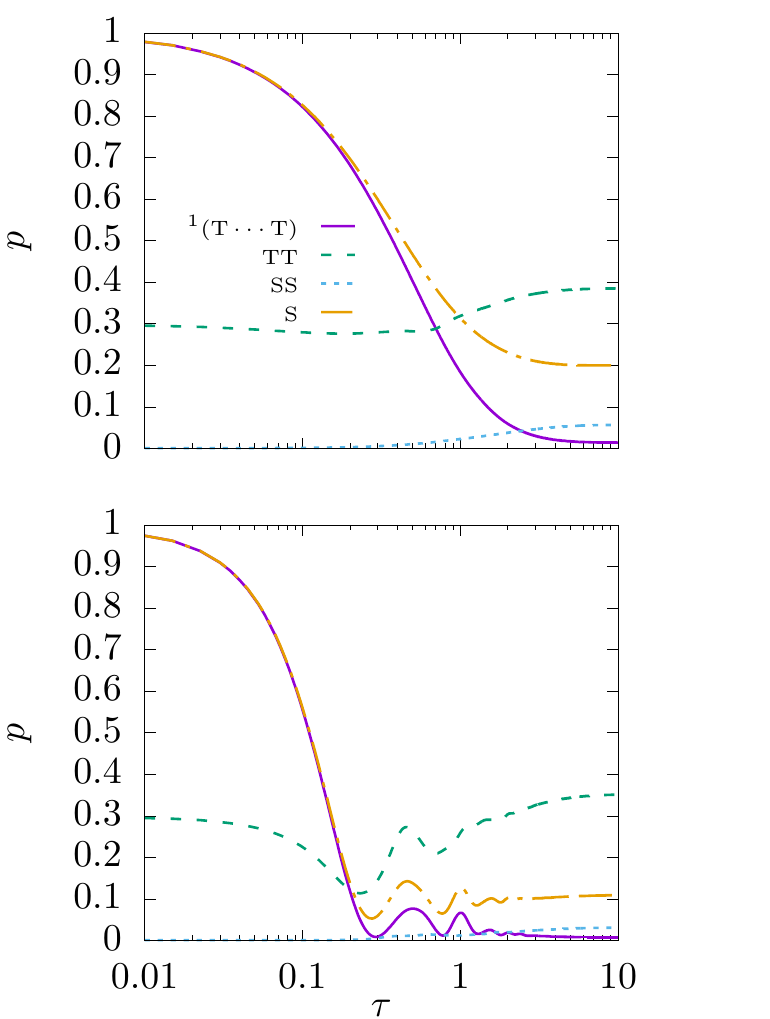}
\caption{\label{Fig:12} As Fig. \ref{Fig:11} but for $N=8$.}
\end{figure}

Fig. \ref{Fig:12} shows the same plots for the larger system of $N=8$. We see similar trends for the case of $A=0$, although the $\TT$ population is further depleted and the population of the TT manifold is stable for short timescales before increasing slightly. The introduction of spin-orbit coupling has a dramatic effect on this picture. The $\TT$ population decreases sharply with $\tau$, while the TT manifold is relatively stable, but showing oscillations around $\tau = 0.5$. The onset and decay of these oscillations corresponds to the increase and then decay of the coherence, as depicted above, highlighting the spin-orbit coupling nature of this effect. A large proportion of population populates high-spin states at the end of the evolution.

\begin{figure}[t]
\centering
\includegraphics[width=0.65\linewidth, clip=true, trim=0 0 30pt 0]{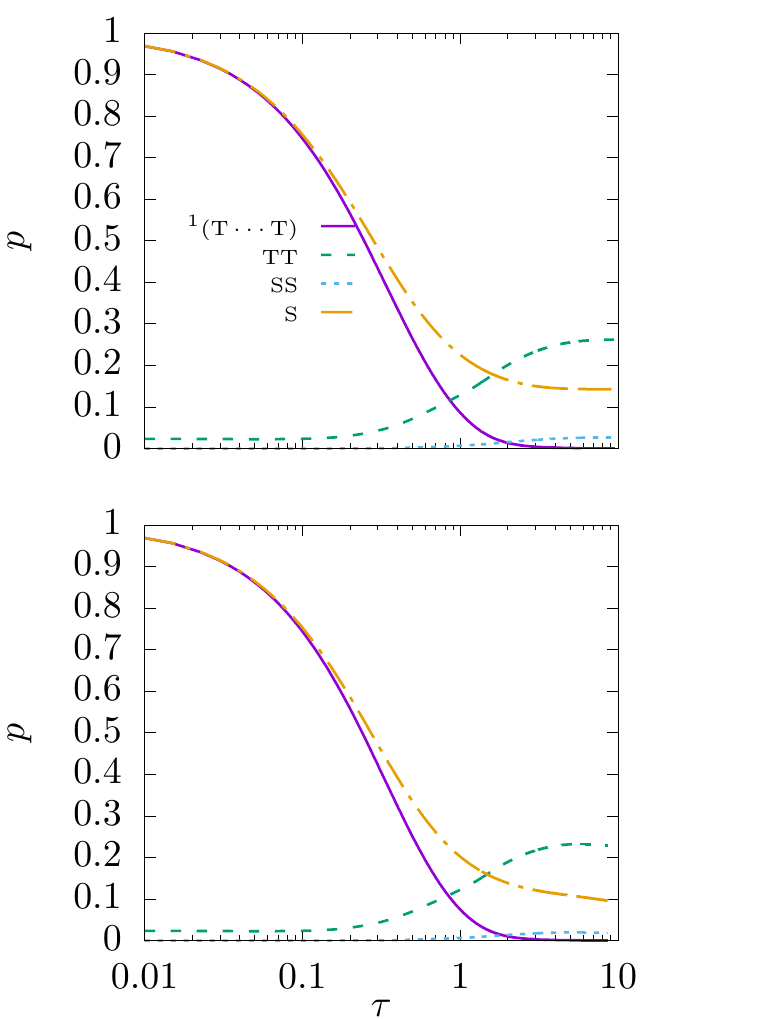}
\caption{\label{Fig:13} As Fig. \ref{Fig:11} but for $N=12$.}
\end{figure}

Finally, Fig. \ref{Fig:13} shows the same for $N=12$. As we have entered the over-damped regime, no oscillations can be seen any more, with all other trends as above. The spin-orbit coupling has a small effect on the long-$\tau$ populations of the TT and S manifolds, both decreasing due to the increased number of states accessible.

\section{Discussion and Outlook}
\subsection{Discussion}
Collecting the information from the numerical simulations we can see the following:
\begin{enumerate}
\item Coherence and entanglement between half-chains decays on the time-scale of $t\simeq \gamma^{-1}$.
\item The total spin and magnetisation increase on the same time scale, $t\simeq \gamma^{-1}$.
\item Under dephasing spin chains will populate high-spin states, which increases the value of $\left\langle S^2\right\rangle$.
\item The population of the $\TT$ state decays into SS states and high spin states.
\item Spin-orbit coupling enhances and accelerates all of the effects with the exception of decoherence, which behaves non-monotonically.
\end{enumerate}

All the results here can be understood in terms of the steady states of the Lindblad master-equation. As justified in Section III, all steady states attainable are diagonal, \ie\ we achieve a statistical mixture of all states that conserve certain quantities. Without spin-orbit coupling these are all states with $M_S = 0$. The dephasing will lead to an equal population of all of these states, leading to an increase of $\left\langle S^2\right\rangle$ with a value of $\hbar^2$ per four sites in the chain. Simultaneously, all off-diagonal elements of the density matrix decay over time, due to damping induced by the environmental interaction. This leads to a decay in coherence and entanglement within the system.

The introduction of spin-orbit coupling in the system lifts the restriction of states with $M_S = 0$. The coupling itself can link states in steps of $M_S = \pm 2$, leading to a larger number of high-spin states becoming available, and therefore enhancing all the processes described above. Spin-orbit coupling also has a non-monotonic effect on observables, namely on $\left\langle S^z\right\rangle$ and the population of states and manifolds, as well as coherence. This non-monotonicity, induced by the same effect, implies some functional link between the observables and the coherence measure. Linking this proxy to the quantum effect measure will be the basis for future research. 

The rapid decay of coherence and entanglement accompanied with depletion of low-spin state population on the same time-scale of $t = \gamma^{-1}$ allows us to theoretically define triplet-triplet decoherence in singlet fission.

In this work we have modelled the interaction of the system (in this case, representing a single polyene chain) with its environment via the Lindblad master equation. To model transverse relaxation (\ie, a dephasing process) we have taken the Lindblad operator to be $\spin{z}{m}/\hbar$. As shown by Eq. (\ref{Eq:23}), this formulation leads to a realistic equation of motion for the relevant observable, whereby it undergoes damping as a consequence of the system-environment coupling. However, an unphysical consequence of this choice of dissipator is that it leads to detailed balance with an effectively infinite temperature. Thus, the system's steady state is an infinite temperature equilibrium distribution over the accessible states. We note, however, that decoherence and the loss of entanglement occur on timescales $\sim T_2$, i.e., before equilibrium is established. Thus, we anticipate that our description of these processes remains qualitatively correct for any kind of environmental interaction.

\subsection{Summary and Outlook}
We have presented numerical and theoretical studies on a uniform Heisenberg spin chain to model singlet fission in carotenoids by interactions with a dephasing environment, taking into account spin-orbit coupling. Both interactions lead to a loss of coherence and entanglement between half-chains. Simultaneously, an increase of the total spin on the same timescale can be observed and in the presence of spin-orbit coupling also an increase of total magnetisation. All these effects show that for efficient singlet-fission in a spin chain, a strong environmental dephasing interaction is beneficial.

Future work will extend the model to larger systems by using different techniques to solve the model Hamiltonian, such as density matrix renormalization group techniques. The introduction of disorder into the model will also be investigated, as well as the coupling to nuclear motion of the underlying atomic framework, and the interaction between multiple chains. In parallel, the theory linking observables to quantum measures will be developed further.

Other forms of the dissipator in the Lindbald formalism or a different form of master equation will also be investigated. By adapting the Lindblad operator it may be possible to establish detailed balance at finite temperature, making our long-time simulations more realistic. Similarly, changing the master equation (for instance using a Redfield equation) may allow for a more physical long-time behaviour.

The influence of external magnetic fields on the effects discussed here will also be investigated. Understanding how observables and the degree of entanglement and coherence between half-chains change under the influence of external fields will allow for the design of novel spectroscopic probes of singlet fission.

\begin{acknowledgments}
The authors would like to thank Jenny Clark and her group, Simon Benjamin and his group, as well as Adam Nahum for useful discussions. We also gratefully acknowledge funding via the EPSRC grant Ref. EP/S002766/1.
\end{acknowledgments}


\bibliography{library.bib}

\begin{thebibliography}{27}%
\makeatletter
\providecommand \@ifxundefined [1]{%
 \@ifx{#1\undefined}
}%
\providecommand \@ifnum [1]{%
 \ifnum #1\expandafter \@firstoftwo
 \else \expandafter \@secondoftwo
 \fi
}%
\providecommand \@ifx [1]{%
 \ifx #1\expandafter \@firstoftwo
 \else \expandafter \@secondoftwo
 \fi
}%
\providecommand \natexlab [1]{#1}%
\providecommand \enquote  [1]{``#1''}%
\providecommand \bibnamefont  [1]{#1}%
\providecommand \bibfnamefont [1]{#1}%
\providecommand \citenamefont [1]{#1}%
\providecommand \href@noop [0]{\@secondoftwo}%
\providecommand \href [0]{\begingroup \@sanitize@url \@href}%
\providecommand \@href[1]{\@@startlink{#1}\@@href}%
\providecommand \@@href[1]{\endgroup#1\@@endlink}%
\providecommand \@sanitize@url [0]{\catcode `\\12\catcode `\$12\catcode
  `\&12\catcode `\#12\catcode `\^12\catcode `\_12\catcode `\%12\relax}%
\providecommand \@@startlink[1]{}%
\providecommand \@@endlink[0]{}%
\providecommand \url  [0]{\begingroup\@sanitize@url \@url }%
\providecommand \@url [1]{\endgroup\@href {#1}{\urlprefix }}%
\providecommand \urlprefix  [0]{URL }%
\providecommand \Eprint [0]{\href }%
\providecommand \doibase [0]{http://dx.doi.org/}%
\providecommand \selectlanguage [0]{\@gobble}%
\providecommand \bibinfo  [0]{\@secondoftwo}%
\providecommand \bibfield  [0]{\@secondoftwo}%
\providecommand \translation [1]{[#1]}%
\providecommand \BibitemOpen [0]{}%
\providecommand \bibitemStop [0]{}%
\providecommand \bibitemNoStop [0]{.\EOS\space}%
\providecommand \EOS [0]{\spacefactor3000\relax}%
\providecommand \BibitemShut  [1]{\csname bibitem#1\endcsname}%
\let\auto@bib@innerbib\@empty
\bibitem [{\citenamefont {Smith}\ and\ \citenamefont
  {Michl}(2010)}]{Smith:2010aa}%
  \BibitemOpen
  \bibfield  {author} {\bibinfo {author} {\bibfnamefont {Millicent~B.}\
  \bibnamefont {Smith}}\ and\ \bibinfo {author} {\bibfnamefont {Josef}\
  \bibnamefont {Michl}},\ }\bibfield  {title} {\enquote {\bibinfo {title}
  {Singlet fission},}\ }\href {\doibase 10.1021/cr1002613} {\bibfield
  {journal} {\bibinfo  {journal} {Chemical Reviews}\ }\textbf {\bibinfo
  {volume} {110}},\ \bibinfo {pages} {6891--6936} (\bibinfo {year}
  {2010})}\BibitemShut {NoStop}%
\bibitem [{\citenamefont {Jadhav}\ \emph {et~al.}(2012)\citenamefont {Jadhav},
  \citenamefont {Brown}, \citenamefont {Thompson}, \citenamefont {Wunsch},
  \citenamefont {Mohanty}, \citenamefont {Yost}, \citenamefont {Hontz},
  \citenamefont {Van~Voorhis}, \citenamefont {Bawendi}, \citenamefont
  {Bulovi{\'c}},\ and\ \citenamefont {Baldo}}]{Jadhav:2012aa}%
  \BibitemOpen
  \bibfield  {author} {\bibinfo {author} {\bibfnamefont {Priya~J.}\
  \bibnamefont {Jadhav}}, \bibinfo {author} {\bibfnamefont {Patrick~R.}\
  \bibnamefont {Brown}}, \bibinfo {author} {\bibfnamefont {Nicholas}\
  \bibnamefont {Thompson}}, \bibinfo {author} {\bibfnamefont {Benjamin}\
  \bibnamefont {Wunsch}}, \bibinfo {author} {\bibfnamefont {Aseema}\
  \bibnamefont {Mohanty}}, \bibinfo {author} {\bibfnamefont {Shane~R.}\
  \bibnamefont {Yost}}, \bibinfo {author} {\bibfnamefont {Eric}\ \bibnamefont
  {Hontz}}, \bibinfo {author} {\bibfnamefont {Troy}\ \bibnamefont
  {Van~Voorhis}}, \bibinfo {author} {\bibfnamefont {Moungi~G.}\ \bibnamefont
  {Bawendi}}, \bibinfo {author} {\bibfnamefont {Vladimir}\ \bibnamefont
  {Bulovi{\'c}}}, \ and\ \bibinfo {author} {\bibfnamefont {Marc~A.}\
  \bibnamefont {Baldo}},\ }\bibfield  {title} {\enquote {\bibinfo {title}
  {Triplet exciton dissociation in singlet exciton fission photovoltaics},}\
  }\href {\doibase 10.1002/adma.201202397} {\bibfield  {journal} {\bibinfo
  {journal} {Advanced Materials}\ }\textbf {\bibinfo {volume} {24}},\ \bibinfo
  {pages} {6169--6174} (\bibinfo {year} {2012})}\BibitemShut {NoStop}%
\bibitem [{\citenamefont {Congreve}\ \emph {et~al.}(2013)\citenamefont
  {Congreve}, \citenamefont {Lee}, \citenamefont {Thompson}, \citenamefont
  {Hontz}, \citenamefont {Yost}, \citenamefont {Reusswig}, \citenamefont
  {Bahlke}, \citenamefont {Reineke}, \citenamefont {Van~Voorhis},\ and\
  \citenamefont {Baldo}}]{Congreve:2013aa}%
  \BibitemOpen
  \bibfield  {author} {\bibinfo {author} {\bibfnamefont {Daniel~N.}\
  \bibnamefont {Congreve}}, \bibinfo {author} {\bibfnamefont {Jiye}\
  \bibnamefont {Lee}}, \bibinfo {author} {\bibfnamefont {Nicholas~J.}\
  \bibnamefont {Thompson}}, \bibinfo {author} {\bibfnamefont {Eric}\
  \bibnamefont {Hontz}}, \bibinfo {author} {\bibfnamefont {Shane~R.}\
  \bibnamefont {Yost}}, \bibinfo {author} {\bibfnamefont {Philip~D.}\
  \bibnamefont {Reusswig}}, \bibinfo {author} {\bibfnamefont {Matthias~E.}\
  \bibnamefont {Bahlke}}, \bibinfo {author} {\bibfnamefont {Sebastian}\
  \bibnamefont {Reineke}}, \bibinfo {author} {\bibfnamefont {Troy}\
  \bibnamefont {Van~Voorhis}}, \ and\ \bibinfo {author} {\bibfnamefont
  {Marc~A.}\ \bibnamefont {Baldo}},\ }\bibfield  {title} {\enquote {\bibinfo
  {title} {External quantum efficiency above 100
  singlet-exciton-fission--based organic photovoltaic cell},}\ }\href
  {https://science.sciencemag.org/content/sci/340/6130/334.full.pdf} {\bibfield
   {journal} {\bibinfo  {journal} {Science}\ }\textbf {\bibinfo {volume}
  {340}},\ \bibinfo {pages} {334--337} (\bibinfo {year} {2013})}\BibitemShut
  {NoStop}%
\bibitem [{\citenamefont {Musser}\ and\ \citenamefont
  {Clark}(2019)}]{Musser:2019aa}%
  \BibitemOpen
  \bibfield  {author} {\bibinfo {author} {\bibfnamefont {Andrew~J.}\
  \bibnamefont {Musser}}\ and\ \bibinfo {author} {\bibfnamefont {Jenny}\
  \bibnamefont {Clark}},\ }\bibfield  {title} {\enquote {\bibinfo {title}
  {Triplet-pair states in organic semiconductors},}\ }\href {\doibase
  10.1146/annurev-physchem-042018-052435} {\bibfield  {journal} {\bibinfo
  {journal} {Annual Review of Physical Chemistry}\ }\textbf {\bibinfo {volume}
  {70}},\ \bibinfo {pages} {323--351} (\bibinfo {year} {2019})}\BibitemShut
  {NoStop}%
\bibitem [{\citenamefont {Nelson}\ \emph {et~al.}(2013)\citenamefont {Nelson},
  \citenamefont {Monahan},\ and\ \citenamefont {Zhu}}]{Nelson:2013aa}%
  \BibitemOpen
  \bibfield  {author} {\bibinfo {author} {\bibfnamefont {Cory~A.}\ \bibnamefont
  {Nelson}}, \bibinfo {author} {\bibfnamefont {Nicholas~R.}\ \bibnamefont
  {Monahan}}, \ and\ \bibinfo {author} {\bibfnamefont {X.-Y.}\ \bibnamefont
  {Zhu}},\ }\bibfield  {title} {\enquote {\bibinfo {title} {Exceeding the
  shockley-queisser limit in solar energy conversion},}\ }\href {\doibase
  10.1039/C3EE42098A} {\bibfield  {journal} {\bibinfo  {journal} {Energy
  Environ. Sci.}\ }\textbf {\bibinfo {volume} {6}},\ \bibinfo {pages}
  {3508--3519} (\bibinfo {year} {2013})}\BibitemShut {NoStop}%
\bibitem [{\citenamefont {Rao}\ and\ \citenamefont
  {Friend}(2017)}]{Rao:2017aa}%
  \BibitemOpen
  \bibfield  {author} {\bibinfo {author} {\bibfnamefont {Akshay}\ \bibnamefont
  {Rao}}\ and\ \bibinfo {author} {\bibfnamefont {Richard~H.}\ \bibnamefont
  {Friend}},\ }\bibfield  {title} {\enquote {\bibinfo {title} {Harnessing
  singlet exciton fission to break the shockley--queisser limit},}\ }\href
  {\doibase 10.1038/natrevmats.2017.63} {\bibfield  {journal} {\bibinfo
  {journal} {Nature Reviews Materials}\ }\textbf {\bibinfo {volume} {2}},\
  \bibinfo {pages} {17063} (\bibinfo {year} {2017})}\BibitemShut {NoStop}%
\bibitem [{\citenamefont {Peterman}\ \emph {et~al.}(1995)\citenamefont
  {Peterman}, \citenamefont {Dukker}, \citenamefont {van Grondelle},\ and\
  \citenamefont {van Amerongen}}]{Peterman:1995aa}%
  \BibitemOpen
  \bibfield  {author} {\bibinfo {author} {\bibfnamefont {E.~J.}\ \bibnamefont
  {Peterman}}, \bibinfo {author} {\bibfnamefont {F.~M.}\ \bibnamefont
  {Dukker}}, \bibinfo {author} {\bibfnamefont {R.}~\bibnamefont {van
  Grondelle}}, \ and\ \bibinfo {author} {\bibfnamefont {H.}~\bibnamefont {van
  Amerongen}},\ }\bibfield  {title} {\enquote {\bibinfo {title} {Chlorophyll a
  and carotenoid triplet states in light-harvesting complex ii of higher
  plants},}\ }\href {\doibase https://doi.org/10.1016/S0006-3495(95)80138-4}
  {\bibfield  {journal} {\bibinfo  {journal} {Biophysical Journal}\ }\textbf
  {\bibinfo {volume} {69}},\ \bibinfo {pages} {2670--2678} (\bibinfo {year}
  {1995})}\BibitemShut {NoStop}%
\bibitem [{\citenamefont {Ritz}\ \emph {et~al.}(2000)\citenamefont {Ritz},
  \citenamefont {Damjanovi{\'c}}, \citenamefont {Schulten}, \citenamefont
  {Zhang},\ and\ \citenamefont {Koyama}}]{Ritz:2000aa}%
  \BibitemOpen
  \bibfield  {author} {\bibinfo {author} {\bibfnamefont {Thorsten}\
  \bibnamefont {Ritz}}, \bibinfo {author} {\bibfnamefont {Ana}\ \bibnamefont
  {Damjanovi{\'c}}}, \bibinfo {author} {\bibfnamefont {Klaus}\ \bibnamefont
  {Schulten}}, \bibinfo {author} {\bibfnamefont {Jian-Ping}\ \bibnamefont
  {Zhang}}, \ and\ \bibinfo {author} {\bibfnamefont {Yasushi}\ \bibnamefont
  {Koyama}},\ }\bibfield  {title} {\enquote {\bibinfo {title} {Efficient light
  harvesting through carotenoids},}\ }\href {\doibase 10.1023/A:1010750332320}
  {\bibfield  {journal} {\bibinfo  {journal} {Photosynthesis Research}\
  }\textbf {\bibinfo {volume} {66}},\ \bibinfo {pages} {125--144} (\bibinfo
  {year} {2000})}\BibitemShut {NoStop}%
\bibitem [{\citenamefont {Pol{\'\i}vka}\ and\ \citenamefont
  {Sundstr{\"o}m}(2004)}]{Polivka:2004aa}%
  \BibitemOpen
  \bibfield  {author} {\bibinfo {author} {\bibfnamefont {Tom{\'a}{\v s}}\
  \bibnamefont {Pol{\'\i}vka}}\ and\ \bibinfo {author} {\bibfnamefont {Villy}\
  \bibnamefont {Sundstr{\"o}m}},\ }\bibfield  {title} {\enquote {\bibinfo
  {title} {Ultrafast dynamics of carotenoid excited states - from solution to
  natural and artificial systems},}\ }\bibfield  {booktitle} {\emph {\bibinfo
  {booktitle} {Chemical Reviews}},\ }\href {\doibase 10.1021/cr020674n}
  {\bibfield  {journal} {\bibinfo  {journal} {Chemical Reviews}\ }\textbf
  {\bibinfo {volume} {104}},\ \bibinfo {pages} {2021--2072} (\bibinfo {year}
  {2004})}\BibitemShut {NoStop}%
\bibitem [{\citenamefont {Pol{\'\i}vka}\ and\ \citenamefont
  {Frank}(2010)}]{Polivka:2010aa}%
  \BibitemOpen
  \bibfield  {author} {\bibinfo {author} {\bibfnamefont {Tom{\'a}{\v s}}\
  \bibnamefont {Pol{\'\i}vka}}\ and\ \bibinfo {author} {\bibfnamefont
  {Harry~A.}\ \bibnamefont {Frank}},\ }\bibfield  {title} {\enquote {\bibinfo
  {title} {Molecular factors controlling photosynthetic light harvesting by
  carotenoids},}\ }\href {\doibase 10.1021/ar100030m} {\bibfield  {journal}
  {\bibinfo  {journal} {Accounts of Chemical Research}\ }\textbf {\bibinfo
  {volume} {43}},\ \bibinfo {pages} {1125--1134} (\bibinfo {year}
  {2010})}\BibitemShut {NoStop}%
\bibitem [{\citenamefont {Casanova}(2018)}]{Casanova:2018aa}%
  \BibitemOpen
  \bibfield  {author} {\bibinfo {author} {\bibfnamefont {David}\ \bibnamefont
  {Casanova}},\ }\bibfield  {title} {\enquote {\bibinfo {title} {Theoretical
  modeling of singlet fission},}\ }\href {\doibase 10.1021/acs.chemrev.7b00601}
  {\bibfield  {journal} {\bibinfo  {journal} {Chemical Reviews}\ }\textbf
  {\bibinfo {volume} {118}},\ \bibinfo {pages} {7164--7207} (\bibinfo {year}
  {2018})}\BibitemShut {NoStop}%
\bibitem [{\citenamefont {Scholes}(2015)}]{Scholes:2015aa}%
  \BibitemOpen
  \bibfield  {author} {\bibinfo {author} {\bibfnamefont {Gregory~D.}\
  \bibnamefont {Scholes}},\ }\bibfield  {title} {\enquote {\bibinfo {title}
  {Correlated pair states formed by singlet fission and exciton--exciton
  annihilation},}\ }\href {\doibase 10.1021/acs.jpca.5b09725} {\bibfield
  {journal} {\bibinfo  {journal} {The Journal of Physical Chemistry A}\
  }\textbf {\bibinfo {volume} {119}},\ \bibinfo {pages} {12699--12705}
  (\bibinfo {year} {2015})}\BibitemShut {NoStop}%
\bibitem [{\citenamefont {Miyata}\ \emph {et~al.}(2019)\citenamefont {Miyata},
  \citenamefont {Conrad-Burton}, \citenamefont {Geyer},\ and\ \citenamefont
  {Zhu}}]{Miyata:2019aa}%
  \BibitemOpen
  \bibfield  {author} {\bibinfo {author} {\bibfnamefont {Kiyoshi}\ \bibnamefont
  {Miyata}}, \bibinfo {author} {\bibfnamefont {Felisa~S.}\ \bibnamefont
  {Conrad-Burton}}, \bibinfo {author} {\bibfnamefont {Florian~L.}\ \bibnamefont
  {Geyer}}, \ and\ \bibinfo {author} {\bibfnamefont {X.~Y.}\ \bibnamefont
  {Zhu}},\ }\bibfield  {title} {\enquote {\bibinfo {title} {Triplet pair states
  in singlet fission},}\ }\href {\doibase 10.1021/acs.chemrev.8b00572}
  {\bibfield  {journal} {\bibinfo  {journal} {Chemical Reviews}\ }\textbf
  {\bibinfo {volume} {119}},\ \bibinfo {pages} {4261--4292} (\bibinfo {year}
  {2019})}\BibitemShut {NoStop}%
\bibitem [{\citenamefont {Pandya}\ \emph {et~al.}(2020)\citenamefont {Pandya},
  \citenamefont {Gu}, \citenamefont {Cheminal}, \citenamefont {Chen},
  \citenamefont {Booker}, \citenamefont {Soucek}, \citenamefont {Schott},
  \citenamefont {Legrand}, \citenamefont {Mathevet}, \citenamefont {Greenham},
  \citenamefont {Barisien}, \citenamefont {Musser}, \citenamefont {Chin},\ and\
  \citenamefont {Rao}}]{Pandaya:2020aa}%
  \BibitemOpen
  \bibfield  {author} {\bibinfo {author} {\bibfnamefont {Raj}\ \bibnamefont
  {Pandya}}, \bibinfo {author} {\bibfnamefont {Qifei}\ \bibnamefont {Gu}},
  \bibinfo {author} {\bibfnamefont {Alexandre}\ \bibnamefont {Cheminal}},
  \bibinfo {author} {\bibfnamefont {Richard Y.~S.}\ \bibnamefont {Chen}},
  \bibinfo {author} {\bibfnamefont {Edward~P.}\ \bibnamefont {Booker}},
  \bibinfo {author} {\bibfnamefont {Richard}\ \bibnamefont {Soucek}}, \bibinfo
  {author} {\bibfnamefont {Michel}\ \bibnamefont {Schott}}, \bibinfo {author}
  {\bibfnamefont {Laurent}\ \bibnamefont {Legrand}}, \bibinfo {author}
  {\bibfnamefont {Fabrice}\ \bibnamefont {Mathevet}}, \bibinfo {author}
  {\bibfnamefont {Neil~C.}\ \bibnamefont {Greenham}}, \bibinfo {author}
  {\bibfnamefont {Thierry}\ \bibnamefont {Barisien}}, \bibinfo {author}
  {\bibfnamefont {Andrew~J.}\ \bibnamefont {Musser}}, \bibinfo {author}
  {\bibfnamefont {Alex~W.}\ \bibnamefont {Chin}}, \ and\ \bibinfo {author}
  {\bibfnamefont {Akshay}\ \bibnamefont {Rao}},\ }\href@noop {} {\enquote
  {\bibinfo {title} {Optical projection and spatial separation of spin
  entangled triplet-pairs from the s1 (21ag-) state of pi-conjugated
  systems},}\ } (\bibinfo {year} {2020}),\ \Eprint
  {http://arxiv.org/abs/2002.12465} {arXiv:2002.12465 [physics.chem-ph]}
  \BibitemShut {NoStop}%
\bibitem [{\citenamefont {Musser}\ \emph {et~al.}(2019)\citenamefont {Musser},
  \citenamefont {Al-Hashimi}, \citenamefont {Heeney},\ and\ \citenamefont
  {Clark}}]{Musser:2019ab}%
  \BibitemOpen
  \bibfield  {author} {\bibinfo {author} {\bibfnamefont {Andrew~J.}\
  \bibnamefont {Musser}}, \bibinfo {author} {\bibfnamefont {Mohammed}\
  \bibnamefont {Al-Hashimi}}, \bibinfo {author} {\bibfnamefont {Martin}\
  \bibnamefont {Heeney}}, \ and\ \bibinfo {author} {\bibfnamefont {Jenny}\
  \bibnamefont {Clark}},\ }\bibfield  {title} {\enquote {\bibinfo {title}
  {Heavy-atom effects on intramolecular singlet fission in a conjugated
  polymer},}\ }\bibfield  {booktitle} {\emph {\bibinfo {booktitle} {The Journal
  of Chemical Physics}},\ }\href {\doibase 10.1063/1.5110269} {\bibfield
  {journal} {\bibinfo  {journal} {The Journal of Chemical Physics}\ }\textbf
  {\bibinfo {volume} {151}},\ \bibinfo {pages} {044902} (\bibinfo {year}
  {2019})}\BibitemShut {NoStop}%
\bibitem [{\citenamefont {Valentine}\ \emph {et~al.}(2020)\citenamefont
  {Valentine}, \citenamefont {Manawadu},\ and\ \citenamefont
  {Barford}}]{Valentine:2020aa}%
  \BibitemOpen
  \bibfield  {author} {\bibinfo {author} {\bibfnamefont {Darren~J.}\
  \bibnamefont {Valentine}}, \bibinfo {author} {\bibfnamefont {Dilhan}\
  \bibnamefont {Manawadu}}, \ and\ \bibinfo {author} {\bibfnamefont {William}\
  \bibnamefont {Barford}},\ }\href@noop {} {\enquote {\bibinfo {title} {Higher
  energy triplet-tiplet singlet states in polyenes and their role in
  intramolecular singlet fission},}\ } (\bibinfo {year} {2020})\BibitemShut
  {NoStop}%
\bibitem [{\citenamefont {Tavan}\ and\ \citenamefont
  {Schulten}(1987)}]{Tavan:1987aa}%
  \BibitemOpen
  \bibfield  {author} {\bibinfo {author} {\bibfnamefont {Paul}\ \bibnamefont
  {Tavan}}\ and\ \bibinfo {author} {\bibfnamefont {Klaus}\ \bibnamefont
  {Schulten}},\ }\bibfield  {title} {\enquote {\bibinfo {title} {Electronic
  excitations in finite and infinite polyenes},}\ }\href {\doibase
  10.1103/PhysRevB.36.4337} {\bibfield  {journal} {\bibinfo  {journal}
  {Physical Review B}\ }\textbf {\bibinfo {volume} {36}},\ \bibinfo {pages}
  {4337--4358} (\bibinfo {year} {1987})}\BibitemShut {NoStop}%
\bibitem [{\citenamefont {Barford}\ \emph {et~al.}(2001)\citenamefont
  {Barford}, \citenamefont {Bursill},\ and\ \citenamefont
  {Lavrentiev}}]{Barford:2001aa}%
  \BibitemOpen
  \bibfield  {author} {\bibinfo {author} {\bibfnamefont {William}\ \bibnamefont
  {Barford}}, \bibinfo {author} {\bibfnamefont {Robert~J.}\ \bibnamefont
  {Bursill}}, \ and\ \bibinfo {author} {\bibfnamefont {Mikhail~Yu}\
  \bibnamefont {Lavrentiev}},\ }\bibfield  {title} {\enquote {\bibinfo {title}
  {Density-matrix renormalization-group calculations of excited states of
  linear polyenes},}\ }\href {\doibase 10.1103/PhysRevB.63.195108} {\bibfield
  {journal} {\bibinfo  {journal} {Physical Review B}\ }\textbf {\bibinfo
  {volume} {63}},\ \bibinfo {pages} {195108--} (\bibinfo {year}
  {2001})}\BibitemShut {NoStop}%
\bibitem [{\citenamefont {Schmidt}\ and\ \citenamefont
  {Tavan}(2012)}]{Schmidt:2012aa}%
  \BibitemOpen
  \bibfield  {author} {\bibinfo {author} {\bibfnamefont {Maximilian}\
  \bibnamefont {Schmidt}}\ and\ \bibinfo {author} {\bibfnamefont {Paul}\
  \bibnamefont {Tavan}},\ }\bibfield  {title} {\enquote {\bibinfo {title}
  {Electronic excitations in long polyenes revisited},}\ }\href {\doibase
  10.1063/1.3696880} {\bibfield  {journal} {\bibinfo  {journal} {The Journal of
  Chemical Physics}\ }\textbf {\bibinfo {volume} {136}},\ \bibinfo {pages}
  {124309} (\bibinfo {year} {2012})}\BibitemShut {NoStop}%
\bibitem [{\citenamefont {Barford}(2013)}]{Barford:2013ab}%
  \BibitemOpen
  \bibfield  {author} {\bibinfo {author} {\bibfnamefont {William}\ \bibnamefont
  {Barford}},\ }\href@noop {} {\emph {\bibinfo {title} {Electronic and Optical
  Properties of Conjugated Polymers}}},\ \bibinfo {edition} {2nd}\ ed.\
  (\bibinfo  {publisher} {Oxford University Press},\ \bibinfo {address}
  {Oxford},\ \bibinfo {year} {2013})\BibitemShut {NoStop}%
\bibitem [{\citenamefont {Barford}\ \emph {et~al.}(2010)\citenamefont
  {Barford}, \citenamefont {Bursill},\ and\ \citenamefont
  {Makhov}}]{Barford:2010aa}%
  \BibitemOpen
  \bibfield  {author} {\bibinfo {author} {\bibfnamefont {William}\ \bibnamefont
  {Barford}}, \bibinfo {author} {\bibfnamefont {Robert~J.}\ \bibnamefont
  {Bursill}}, \ and\ \bibinfo {author} {\bibfnamefont {Dmitry~V.}\ \bibnamefont
  {Makhov}},\ }\bibfield  {title} {\enquote {\bibinfo {title} {Spin-orbit
  interactions between interchain excitations in conjugated polymers},}\ }\href
  {\doibase 10.1103/PhysRevB.81.035206} {\bibfield  {journal} {\bibinfo
  {journal} {Physical Review B}\ }\textbf {\bibinfo {volume} {81}},\ \bibinfo
  {pages} {035206--} (\bibinfo {year} {2010})}\BibitemShut {NoStop}%
\bibitem [{\citenamefont {Alicki}\ and\ \citenamefont
  {Lendi}(2007)}]{Alicki:2007aa}%
  \BibitemOpen
  \bibfield  {author} {\bibinfo {author} {\bibfnamefont {Robert}\ \bibnamefont
  {Alicki}}\ and\ \bibinfo {author} {\bibfnamefont {K.}~\bibnamefont {Lendi}},\
  }\href@noop {} {\emph {\bibinfo {title} {Quantum Dynamical Semigroups and
  Applications}}}\ (\bibinfo  {publisher} {Springer-Verlag},\ \bibinfo
  {address} {Berlin Heidelberg},\ \bibinfo {year} {2007})\BibitemShut {NoStop}%
\bibitem [{\citenamefont {Baumgratz}\ \emph {et~al.}(2014)\citenamefont
  {Baumgratz}, \citenamefont {Cramer},\ and\ \citenamefont
  {Plenio}}]{Baumgratz:2014aa}%
  \BibitemOpen
  \bibfield  {author} {\bibinfo {author} {\bibfnamefont {T.}~\bibnamefont
  {Baumgratz}}, \bibinfo {author} {\bibfnamefont {M.}~\bibnamefont {Cramer}}, \
  and\ \bibinfo {author} {\bibfnamefont {M.~B.}\ \bibnamefont {Plenio}},\
  }\bibfield  {title} {\enquote {\bibinfo {title} {Quantifying coherence},}\
  }\href {\doibase 10.1103/PhysRevLett.113.140401} {\bibfield  {journal}
  {\bibinfo  {journal} {Physical Review Letters}\ }\textbf {\bibinfo {volume}
  {113}},\ \bibinfo {pages} {140401} (\bibinfo {year} {2014})}\BibitemShut
  {NoStop}%
\bibitem [{\citenamefont {Chitambar}\ and\ \citenamefont
  {Gour}(2019)}]{Chitambar:2019aa}%
  \BibitemOpen
  \bibfield  {author} {\bibinfo {author} {\bibfnamefont {Eric}\ \bibnamefont
  {Chitambar}}\ and\ \bibinfo {author} {\bibfnamefont {Gilad}\ \bibnamefont
  {Gour}},\ }\bibfield  {title} {\enquote {\bibinfo {title} {Quantum resource
  theories},}\ }\href {\doibase 10.1103/RevModPhys.91.025001} {\bibfield
  {journal} {\bibinfo  {journal} {Reviews of Modern Physics}\ }\textbf
  {\bibinfo {volume} {91}},\ \bibinfo {pages} {025001--} (\bibinfo {year}
  {2019})}\BibitemShut {NoStop}%
\bibitem [{\citenamefont {Horodecki}\ \emph {et~al.}(2009)\citenamefont
  {Horodecki}, \citenamefont {Horodecki}, \citenamefont {Horodecki},\ and\
  \citenamefont {Horodecki}}]{Horodecki:2009aa}%
  \BibitemOpen
  \bibfield  {author} {\bibinfo {author} {\bibfnamefont {Ryszard}\ \bibnamefont
  {Horodecki}}, \bibinfo {author} {\bibfnamefont {Pawe{\l}}\ \bibnamefont
  {Horodecki}}, \bibinfo {author} {\bibfnamefont {Micha{\l}}\ \bibnamefont
  {Horodecki}}, \ and\ \bibinfo {author} {\bibfnamefont {Karol}\ \bibnamefont
  {Horodecki}},\ }\bibfield  {title} {\enquote {\bibinfo {title} {Quantum
  entanglement},}\ }\href {\doibase 10.1103/RevModPhys.81.865} {\bibfield
  {journal} {\bibinfo  {journal} {Reviews of Modern Physics}\ }\textbf
  {\bibinfo {volume} {81}},\ \bibinfo {pages} {865--942} (\bibinfo {year}
  {2009})}\BibitemShut {NoStop}%
\bibitem [{\citenamefont {Streltsov}\ \emph {et~al.}(2015)\citenamefont
  {Streltsov}, \citenamefont {Singh}, \citenamefont {Dhar}, \citenamefont
  {Bera},\ and\ \citenamefont {Adesso}}]{Streltsov:2015aa}%
  \BibitemOpen
  \bibfield  {author} {\bibinfo {author} {\bibfnamefont {Alexander}\
  \bibnamefont {Streltsov}}, \bibinfo {author} {\bibfnamefont {Uttam}\
  \bibnamefont {Singh}}, \bibinfo {author} {\bibfnamefont {Himadri~Shekhar}\
  \bibnamefont {Dhar}}, \bibinfo {author} {\bibfnamefont {Manabendra~Nath}\
  \bibnamefont {Bera}}, \ and\ \bibinfo {author} {\bibfnamefont {Gerardo}\
  \bibnamefont {Adesso}},\ }\bibfield  {title} {\enquote {\bibinfo {title}
  {Measuring quantum coherence with entanglement},}\ }\href {\doibase
  10.1103/PhysRevLett.115.020403} {\bibfield  {journal} {\bibinfo  {journal}
  {Physical Review Letters}\ }\textbf {\bibinfo {volume} {115}},\ \bibinfo
  {pages} {020403--} (\bibinfo {year} {2015})}\BibitemShut {NoStop}%
\bibitem [{\citenamefont {Zhu}\ \emph {et~al.}(2017)\citenamefont {Zhu},
  \citenamefont {Ma}, \citenamefont {Cao}, \citenamefont {Fei},\ and\
  \citenamefont {Vedral}}]{Zhu:2017ab}%
  \BibitemOpen
  \bibfield  {author} {\bibinfo {author} {\bibfnamefont {Huangjun}\
  \bibnamefont {Zhu}}, \bibinfo {author} {\bibfnamefont {Zhihao}\ \bibnamefont
  {Ma}}, \bibinfo {author} {\bibfnamefont {Zhu}\ \bibnamefont {Cao}}, \bibinfo
  {author} {\bibfnamefont {Shao-Ming}\ \bibnamefont {Fei}}, \ and\ \bibinfo
  {author} {\bibfnamefont {Vlatko}\ \bibnamefont {Vedral}},\ }\bibfield
  {title} {\enquote {\bibinfo {title} {Operational one-to-one mapping between
  coherence and entanglement measures},}\ }\href {\doibase
  10.1103/PhysRevA.96.032316} {\bibfield  {journal} {\bibinfo  {journal}
  {Physical Review A}\ }\textbf {\bibinfo {volume} {96}},\ \bibinfo {pages}
  {032316--} (\bibinfo {year} {2017})}\BibitemShut {NoStop}%
\end{thebibliography}%

\appendix

\section{Spin-Orbit Coupling for the Heisenberg model}
Spin-orbit coupling manifests itself within spin chains as the spin-flip of migrating electrons. We can write \cite{Barford:2010aa},
\begin{equation}
\hat{H}_{\text{SO}} = \tilde{A}\sum_i\sum_{\sigma}\left(\hat{c}_{i\sigma}^{\dagger}\hat{c}_{i+1,\bar{\sigma}} - \hat{c}_{i+1,\sigma}^{\dagger}\hat{c}_{i\bar{\sigma}}\right),
\end{equation}
where $\tilde{A}^* = -\tilde{A}$, \ie\ it is purely imaginary and $\hat{H}_{\text{SO}}$ is hence hermitian. $\hat{c}^{\dagger}_{i\sigma}$ ($\hat{c}_{i\sigma}$) creates (destroys) an excited electron of spin $\sigma$ on site $i$ and $\bar{\sigma}$ is the complementary spin to $\sigma$. We now need to map this behaviour onto the Heisenberg model. For a two-site model we can write the terms explicitly as,
\begin{equation}
\hat{H}_{\text{SO}} = \tilde{A}\left(\hat{c}_{1\uparrow}^{\dagger}\hat{c}_{2\downarrow} - \hat{c}^{\dagger}_{2\uparrow}\hat{c}_{1\downarrow} + \hat{c}_{1\downarrow}^{\dagger}\hat{c}_{2\uparrow}-\hat{c}_{2\downarrow}^{\dagger}\hat{c}_{1\uparrow}\right).
\label{Eq:A2}
\end{equation}
\begin{figure*}
\centering
\includegraphics{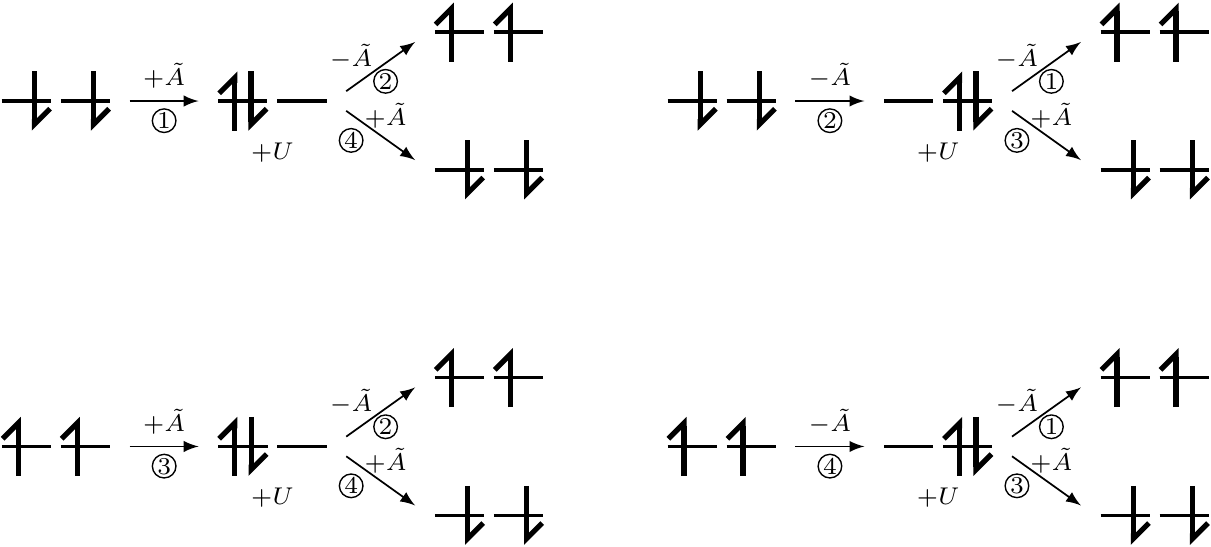}
\caption{\label{Fig:A1} Schematic depicting virtual hopping of a two-site-two-spin model system via the terms of $\hat{H}_{\text{SO}}$ (Eq. (\ref{Eq:A2})) and the associated energy scales.}
\end{figure*}
Using a Schrieffer-Wolff transformation allows us to eliminate any hopping terms to first order. We can express virtual hopping schematically as done in Fig. \ref{Fig:A1}. We see that we have four diagonal terms for parallel spins which contribute $2\tilde{A}^2/U$, while the other four terms flip the spin pairs, contributing $-2\tilde{A}^2/U$. We therefore can express the effective Hamiltonian for $N$ sites as,
\begin{equation}
\hat{H}_{\text{SO}}^{\text{eff}} = \frac{2\tilde{A}^2}{U}\sum_i^N \left(\sum_{\sigma} \hat{n}_{i\sigma}\hat{n}_{i+1,\sigma} - \frac{1}{\hbar^2}\left(\spin{+}{i}\spin{+}{i+1} + \spin{-}{i}\spin{-}{i+1}\right)\right).
\end{equation}
By expressing $\hat{n}_{i\uparrow} = \spin{+}{i}\spin{-}{i}$ and $\hat{n}_{i\downarrow} = \spin{-}{i}\spin{+}{i}$ we have therefore mapped the spin-orbit coupling onto the Heisenberg model. In the main text we have collected the constants in front of the expression and renamed $A$.

\section{Incoherent Operations}
Consider the incoherent state $\hat{\rho} = N^{-1} \sum_{i=1}^N \ket{i}\bra{i}\subset\mathbb{I}$, where $\mathbb{I}$ is the set of incoherent states and the basis states are eigenstates of the Heisenberg Hamiltonian. We can show that the operations $\hat{L}_m = \spin{+}{m}/\hbar$ and $\hat{L}_m = \spin{z}{m}/\hbar$ are incoherent operations by showing that the resulting state is also incoherent. It is straightforward to show that $\spin{z}{m}$ is an incoherent operation, \emph{viz.},
\begin{equation}
\frac{1}{N}\sum_{i=1}^N\spin{z}{m}\ket{i}\bra{i}\spin{z}{m} = \frac{\hbar^2}{4N}\sum_{i=1}^N \ket{i}\bra{i}\subset\mathbb{I},
\end{equation}
which is also an incoherent state. For the spin-flip operations we find,
\begin{equation}
\frac{1}{N}\sum_{i=1}^N\spin{+}{m}\ket{i}\bra{i}\spin{-}{m} = \frac{\hbar^2}{4N}\sum_{i=1}^N\ket{i_{m\uparrow}}\bra{i_{m\uparrow}} \subset\mathbb{I},
\end{equation}
where $\ket{i_{m\uparrow}}$ is the state $\ket{i}$ with the spin on site $m$ flipped from $\beta$ to $\alpha$, if possible. Naturally, some of these states will vanish, but that does not impact the generality of the preceding derivation.

\end{document}